\documentclass[aps,prb,reprint,superscriptaddress,showpacs]{revtex4-2}

\usepackage{amsfonts}
\usepackage{amsmath}
\usepackage{graphicx}
\usepackage{hyperref}
\usepackage{dcolumn}
\usepackage{bm}
\usepackage{float}
\usepackage{xcolor}

\begin{document}
	
	\title{Ultrafast laser-stimulated field emission from a metal: a non-equilibrium Green's functions approach}
	
	\author{Luciano Jacopo D'Onofrio}
	\email{lucianojacopo.donofrio@spin.cnr.it}
	\affiliation{Dipartimento di Fisica ``E.R. Caianiello'', Universit\`{a} degli Studi di Salerno, I-84084 Fisciano (SA), Italy}
	\affiliation{Department of Physics, University of Erlangen-N\"{u}rnberg, 91058 Erlangen, Germany}
	\affiliation{CNR-SPIN, UoS di Salerno, I-84084 Fisciano (SA), Italy}
	\author{Adolfo Avella}
	\affiliation{Dipartimento di Fisica ``E.R. Caianiello'', Universit\`{a} degli Studi di Salerno, I-84084 Fisciano (SA), Italy}
	\affiliation{CNR-SPIN, UoS di Salerno, I-84084 Fisciano (SA), Italy}
	\affiliation{Unit\`{a} CNISM di Salerno, Universit\`{a} degli Studi di Salerno, I-84084 Fisciano (SA), Italy}
	\author{Martin Eckstein}
	\affiliation{I. Institute for Theoretical Physics, University of Hamburg, Notkestra{\ss}e 9-11, 22607 Hamburg, Germany}
	\affiliation{The Hamburg Centre for Ultrafast Imaging, Hamburg, Germany}
	
	\begin{abstract}
		Ultrafast laser-assisted field emission from metallic nanotips is a promising route for generating coherent electron pulses.
		However, existing theoretical models
		often neglect many-body strong correlation effects in the emitter's electronic structure and its response to the optical field.
		In this work, we propose a theoretical framework based on the non-equilibrium Green's function formalism to simulate the out-of-equilibrium electron dynamics across a strongly driven matter-vacuum interface,
		in a one-dimensional model.
		Treating the tip-vacuum coupling perturbatively, we find that the interplay between
		strong-field emission and multi-photon emission regimes
		can be described in terms of a
		filtering of the emitter's local density of occupied states through
		field-dressed vacuum propagators.
		By including electron-electron
		interactions via the second-Born approximation, we show that correlation-induced modifications in the electronic structure are directly transferred to the far-field spectrum.
		Finally, by self-consistently solving the Dyson equation at the interface, we uncover a dynamic back-action regime driven by the optical field, characterized by a periodic polarization of the emitter, which reshapes the far-field spectrum. This shows that the non-equilibrium Green's functions approach provides a versatile framework for exploring and analyzing field emission of correlated electrons.
	\end{abstract}
	
	\maketitle
	
	\section{Introduction}
	
	Field emission is a well-known process in which electrons are emitted from a material when a voltage difference is applied between the material and an electrode. Such a phenomenon has been widely investigated~\cite{PhysRev.102.1464, gomer1961field}, especially after the seminal work by R.H. Fowler and L.W. Nordheim~\cite{fowler1928electron, nordhiem1928effect}, where the one-dimensional time-independent Schr\"{o}dinger equation was solved in the case of an electron tunneling from a bulk metal into the vacuum through a triangular barrier. By adopting their approach, one can infer the Fowler-Nordheim (FN) formula, which relates the current density of the emitted electrons to the material work function and the applied electric field (also known as the barrier field), and
	which can be further adapted for reproducing more realistic steady-state cases~\cite{forbes2007reformulation, dong_quantum_2018}. Recent developments in laser technology~\cite{brabec2000intense,krausz2009attosecond} have paved the way for generating very small sources of free-electron pulses on extremely short timescales by combining field emission with ultrafast light pulses~\cite{PhysRevLett.96.077401, PhysRevLett.97.247402, PhysRevLett.105.147601, PhysRevLett.105.257601, PhysRevLett.98.043907, Kruger2012}. Obtaining electronic pulses with simultaneous high spatial and temporal resolution is a crucial objective
	with compelling, multifaceted applications~\cite{Zhang2016, Zhou2019}
	that include free-electron lasers~\cite{OShea}, picosecond cathodoluminescence~\cite{Merano2005}, time-resolved electron microscopy~\cite{Niikura2002, doi:10.1126/science.1090052}, optical accelerators~\cite{RevModPhys.86.1337}, single-electron sources~\cite{Feve2007},
	and carrier-envelope phase detection~\cite{Kruger2011}.
	
	The setup for an ultrafast laser-assisted electron emission experiment typically consists of a sharp, voltage-biased tip with a tightly focused laser beam at its apex. The voltage difference accelerates the emitted electrons toward an electrode, where they are detected either through a micro-channel plate coupled to a phosphor screen or via an electron spectrometer~\cite{Kruger2012}. Nanotips are particularly useful for investigating laser-stimulated electron emission. Indeed, in light of their geometrical shape, field enhancement due to a plasmonic resonance~\cite{Barnes2003} or
	due to the so-called lightning-rod effect~\cite{AVErmushev_1993} can be efficiently exploited~\cite{thomas2015large}. More broadly, this interplay between nanoscale morphology and field enhancement provides a well-established strategy for governing emission characteristics across various materials, from carbon nanotubes~\cite{deHeer1995, Fan1999} to semiconducting
	oxides~\cite{Lee2002ZnO} and graphene-based
	emitters~\cite{graph1, graph2}.
	
	Achieving clear, overall control over the features of emitted electronic pulses, such as their energy or spatial distribution, requires developing theoretical frameworks that can interpret experimental findings and emission regimes.
	Therefore, over the years, several approaches have been developed. Generalizing the FN formula by defining an effective laser-modified work function or barrier field was among the first attempts to distinguish among the different emission regimes~\cite{PhysRevLett.96.077401}. However, the FN theory is not well suited to describe out-of-equilibrium processes, especially when the duration of the light pulses is comparable to the electronic timescale within the material~\cite{PhysRevLett.97.247402}. For that reason, theoretical efforts were dedicated to integrating the time-dependent Schr\"{o}dinger equation~\cite{PhysRevLett.97.247402, Zhang2016} along a single dimension, in light of the confinement of the electron emission to the tip apex~\cite{Kruger_2012}, also using the time-dependent perturbation theory within the strong-field approximation (SFA) framework~\cite{PhysRevLett.98.043907, PhysRevA.22.1786}. Even though those theories are in good qualitative agreement with the experimental data, they neglect the laser-stimulated electron dynamics within the tip, which is often modeled as a simple quantum well. One possible way to overcome this limitation is to solve the Boltzmann equation for the non-equilibrium single-electron distribution in the matter, accounting for the various mechanisms at play, such as heating from laser excitation or cooling from collisional effects~\cite{PhysRevB.81.115429, PhysRevLett.107.087601}. In doing so, it becomes clear that, for instance,
	the field emission is not only determined by electrons close to the Fermi level~\cite{PhysRevB.88.195434},
	and the electronic structure over a wider range, including possibly correlation effects, should be considered. One must therefore understand how pulse penetration into the material can dynamically affect its electronic properties, and whether laser-stimulated electron emission leads to a back-action effect on the material itself and, in turn, contributes to the emission process.
	
	Moreover, established models for investigating laser-assisted field emission are
	often
	founded on the single-particle assumption.
	This is well justified
	after electrons have been
	emitted into the vacuum, but can become inaccurate when the tip contains strongly correlated electrons, whose light-driven dynamics has unveiled a rich variety of phenomena~\cite{giannetti_capone_fausti_16, delaTorre2021, Murakami2025}.
	An understanding of how the effects of the interactions manifest as measurable quantities in the field emission would be interesting from a theoretical perspective, and could also pave the way to use ultrafast laser-assisted emission techniques to investigate the properties of materials.
	A complete description of field emission from such a system would require treating the non-equilibrium dynamics of the correlated material and the subsequent emission and propagation of electrons within a common framework. Conceptually, this resembles the one-step model of photoemission~\cite{PhysRevB.15.1728,PENDRY1976679,HOPKINSON198069}, in which excitation and photoelectron propagation are treated as a single coherent process. For time-resolved photoemission from laser-driven solids, related approaches have been formulated using non-equilibrium Green's functions (NEGFs) to describe the interplay between Floquet-dressed initial states and Volkov-dressed final states~\cite{Sch_ler_2021, h59r-hyc4}. Extending these calculations to an explicitly spatially inhomogeneous geometry, in which the correlated material, the matter-vacuum interface, and the emitted electrons are all part of the dynamical many-body problem, however, is computationally demanding.
	Here we instead propose
	a minimal setup for reproducing the extraction of electrons from a voltage-biased metallic nanotip driven by an ultrarapid electromagnetic pulse acting at the interface between the tip and the vacuum. By implementing it, we can infer the time-integrated energy distribution of the laser-emitted electrons far from the interface (the far-field energy distribution). To account for the emitter properties, we model the tip using a single-band, one-dimensional tight-binding chain. The out-of-equilibrium electron dynamics along the inhomogeneous (matter chain + vacuum medium) system is described within the non-equilibrium Green's function framework~\cite{RevModPhys.86.779, Kamenev:2011, Stefanucci-Leeuwen:2013, Eckstein2018}, and simulated through the implementation of the NonEquilibrium Systems Simulation Library (NESSi)~\cite{SCHULER2020107484}. By choosing local self-energy terms, we can ``disentangle'' the two chains when solving the equations of motion. In this way, we interpret the far-field energy distribution as the result of filtering an emitter term through isolated vacuum propagators.
	
	This paper is organized as follows. In Sec.~\ref{sec:Model-and-Method}, we outline the physical model and the methodology, introducing the tight-binding Hamiltonians for the matter and vacuum regions, the NEGF formalism, and the algorithm employed to evaluate the far-field energy distribution. The analytical framework for the perturbative tunneling regime is also discussed. In Sec.~\ref{sec:Results}, we present and discuss our numerical results. We first define the time-dependent vacuum barrier profile and benchmark our approach within the uncorrelated case, mapping the different emission regimes in terms of the field features. Then, we consider the weakly correlated regime to investigate how electron-electron interactions reshape the local density of occupied states and dress the emitted quasiparticles. Finally, we go beyond the perturbative limit to analyze the laser pulse's dynamic back-action on the emitting tip. Our findings and future perspectives are summarized in Sec.~\ref{sec:Summary-and-Conclusions}.
	
	\section{Model and Methodology\label{sec:Model-and-Method}}
	
	\subsection{Hamiltonian}
	
	We consider a metallic tip in equilibrium at a temperature $T$ in the distant past ($t\rightarrow-\infty$). We assume that electrons are confined inside the metallic tip at $t\rightarrow-\infty$ because the metal work function $\phi$ is sufficiently high such that $e^{-\beta \phi}\ll1$, with $\beta=1/k_{B}T$. At time $t=t_{0}=0$, a voltage bias $V_{pot}$ is applied between the metallic tip and an electrode, bending the work-function step barrier and thus allowing electrons to tunnel through it. Then, at time $t=t_{1}>t_{0}$, an intense ultrafast laser pulse with angular frequency $\omega_{ph}$ is switched on at the interface between the emitter and the vacuum, affecting the emission process and modifying the potential energy $\mathcal{U}=\mathcal{U}\left(\mathbf{r},t\right)$ of an electron moving across the vacuum. Electrons can, therefore, propagate toward a detector placed in a region of space sufficiently far from the matter-vacuum interface (the far-field region) where their time-resolved energy distribution is measured. In that region, the single-electron energy can be fairly approximated by $\varepsilon_{ff}\left(\mathbf{k}\right)=k^{2}+\phi-V_{pot}$, where $\mathbf{k}$ is the electronic momentum, and $\hbar=2m=e=1$ from here on. Here, we show how the electronic far-field energy distribution can be theoretically estimated using a tight-binding approach within the NEGF formalism.
	
	We assume that the tip is sufficiently thin to allow for a one-dimensional treatment along the normal direction, $x$, to the surface of the metallic tip, with the matter-vacuum interface located at $x=0$. Thus, the Hamiltonian of the whole system, $\hat{\mathcal{H}}$, can be expressed as the sum of three terms: one corresponding to the tip ($x<0$), $\hat{\mathcal{H}}_m$, one describing the vacuum ($x>0$), $\hat{\mathcal{H}}_v$, and a third term representing the coupling between the tip and the vacuum, $\hat{\mathcal{H}}_{c}$. To model the metallic tip, we adopt
	a single-orbital one-dimensional Hubbard Hamiltonian,
	\begin{align}
		\hat{\mathcal{H}}_{m}(t) & =-t_{m}\sum_{\sigma}\sum_{i=1}^{N-1}\left(\hat{c}_{i\sigma}^{\dagger}\hat{c}_{i+1\sigma}+h.c.\right)+\nonumber \\
		+ & \sum_{\sigma}\sum_{i=1}^{N}\left[\epsilon_{m}(t)-\frac{U}{2}\right]\hat{c}_{i\sigma}^{\dagger}\hat{c}_{i\sigma}+U\sum_{i=1}^{N}\hat{c}_{i\uparrow}^{\dagger}\hat{c}_{i\uparrow}\hat{c}_{i\downarrow}^{\dagger}\hat{c}_{i\downarrow},
	\end{align}
	where $N$ is the number of sites along the chain, $t_{m}$ is the hopping amplitude (chosen real without any loss of generality) between two nearest-neighbor sites, $\epsilon_{m}$ corresponds to the on-site mean energy potential, and $U$ is the local Coulomb interaction. $\hat{c}_{i\sigma}^{\dagger}$ ($\hat{c}_{i\sigma}$) creates (annihilates) an electron on the $i$-th site of the chain with spin $\sigma=\uparrow,\downarrow$. Here, we fix the chemical potential $\mu$ at 0, and consider the metallic chain prepared in the half-filling condition at equilibrium: $\langle \hat{c}_{i\sigma}^{\dagger}\hat{c}_{i\sigma}\rangle \left(t \rightarrow -\infty \right) = \langle n_{i\sigma} \rangle \left(t \rightarrow -\infty \right) = 0.5$ for $i=1,\ldots, N$ and $\sigma=\uparrow,\downarrow$.
	Furthermore, the Hamiltonian describing the vacuum medium is given by
	\begin{equation}
		\hat{\mathcal{H}}_{v}\left(t\right)=\sum_{\sigma}\int_{0}^{\infty}dx\,\hat{\psi}_{\sigma}^{\dagger}\left(x\right)\left[-\partial_{x}^{2}+\mathcal{U}\left(x,t\right)\right]\hat{\psi}_{\sigma}\left(x\right),
	\end{equation}
	where $\hat{\psi}_{\sigma}$ is the quantum field operator for an electron with spin $\sigma$ propagating for $x>0$. Finally, we assume that electrons can leave the matter chain only by hopping from the last site of the chain into the vacuum. Thus, the Hamiltonian term for the matter-vacuum coupling can be written as
	\begin{equation}
		\hat{\mathcal{H}}_{c}\left(t\right)=\sum_{\sigma}\int_{0}^{\infty}dx\,\left[J^{\sigma}\left(x,t\right)\hat{\psi}^{\dagger}_{\sigma}\left(x\right)\hat{c}_{N\sigma}+h.c.\right],
	\end{equation}
	where $J^{\sigma}=J^{\sigma}(x,t)$ is the single-particle current density associated with the probability that an electron with spin $\sigma$ hops from the last site $N$ of the chain to a generic point $x>0$ in the vacuum at time $t$.
	
	\subsection{NEGF formalism}
	To describe and simulate the electron dynamics during and after the laser pulse, we adopt the NEGF formalism. Starting with the matter chain, we consider the space- and time-dependent Green's functions $G_{i,j}\left(t,t'\right)=-i\left\langle \mathcal{T}_{\mathcal{C}}\hat{c}_{i}\left(t\right)\hat{c}_{j}^{\dagger}\left(t'\right)\right\rangle $, where we discard the spin index $\sigma$ from here on, $t$ and $t'$ lie on the $L$-shaped Kadanoff-Baym (KB) contour $\mathcal{C}$, and $\mathcal{T}_{\mathcal{C}}$ is the contour-ordering operator. Then, we assume a local self-energy, $\Sigma_{i,j}\left(t,t'\right)=\Sigma_{i}\left(t,t'\right)\delta_{i,j}$.
	This assumption corresponds to the dynamical mean-field theory (DMFT) approximation and, thus, captures the main effects of local electronic correlations, including Mott physics at large $U$, and electron-electron scattering and thermalization in the metallic phase~\cite{RevModPhys.86.779}. Furthermore, it reduces the Dyson equation for the full two-time matrix $G_{i,j}(t,t')$ to a set of recursive equations for the local Green's functions $G_{i,i}(t,t')$ alone (the layer DMFT)~\cite{Potthoff_PhysRevB.59.2549, PhysRevB.70.195342}. Working instead with the full $N\times N$ two-time Green's function would be numerically too costly.
	The derivation of the layer DMFT equations on the Keldysh contour is found in Ref.~\cite{PhysRevB.88.075135};
	here we will only summarize the resulting equations. The two-time diagonal elements $G_{i,i}\left(t,t'\right)$ can be obtained starting from~\cite{PhysRevB.88.075135}
	\begin{equation}
		G_{i,i}=\frac{1}{\mathcal{G}_{i}^{-1}-\Delta_{i-1}^{l}-\Delta_{i+1}^{r}},\label{eq:dys_eq_trid}
	\end{equation}
	with $\mathcal{G}_{i}^{-1}=i\partial_{t}-\epsilon_{i}\left(t\right)-{\Sigma}_{i,i}$, $\Delta_{i-1}^{l}\left(t,t'\right)=t_{i-1}^{*}\left(t\right){G}_{i-1,i-1}^{\left[i\right]}\left(t,t'\right)t_{i-1}\left(t'\right)$, and $\Delta_{i+1}^{r}\left(t,t'\right)=t_{i}\left(t\right){G}_{i+1,i+1}^{\left[i\right]}\left(t,t'\right)t_{i}^{*}\left(t'\right)$. $G_{i,i}^{\left[j\right]}$ stands for the $\left(i,i\right)$ element of the Green's function matrix associated with the chain where the $j$-th site is removed. It can be shown that the term
	$\Delta_{N+1}^{r}$, i.e., the correction to the self-energy on the last matter site due to the coupling with the vacuum, corresponds to
	\begin{equation}
		\Delta\left(t,t'\right)=\int_{0}^{\infty}dxdx'J\left(X\right)g\left(X,X'\right)J^{*}\left(X'\right).
	\end{equation}
	Here, $X=\left(x,t\right)$ and $g=g\left(X, X'\right)$ is the NEGF for an electron propagating in the vacuum when it is ``disentangled'' from the matter chain. On the other hand, it can be
	proven that $\xi \left(X,X'\right)=J\left(x,t\right)G_N\left(t,t'\right)J^{*}\left(x',t'\right)$ acts as a self-energy term for the vacuum, where $G_N=G_N\left(t,t'\right)$ is the electronic NEGF on the site $N$ when the tip is isolated. Thus, we can conclude that
	\begin{align}
		\mathbb{G}\left(X,X'\right) & =g\left(X,X'\right)+\int dX_{1}dX_{2}\times\nonumber \\
		\times & \left[g\left(X,X_{1}\right)J\left(X_{1}\right)\mathcal{T}_{N}\left(t_{1},t_{2}\right)J^{*}\left(X_{2}\right)g\left(X_{2},X'\right)\right]\label{eq:full_G_vac}
	\end{align}
	where $\mathbb{G}=\mathbb{G}\left(X,X'\right)$ is the single-electron NEGF when the vacuum is coupled to the emitter,
	the integration is performed over space for $x>0$ and along the $L$-shaped contour, and the two-time
	$T$-matrix $\mathcal{T}_{N}$ is defined as
	\begin{equation}
		\label{aghsjssass}
		\mathcal{T}_{N}=G_{N}+G_{N}\ast\Delta\ast G_{N}+G_{N}\ast\Delta\ast G_{N}\ast\Delta\ast G_{N}+\ldots,
	\end{equation}
	which corresponds to the Dyson equation $\left(1-G_{N}\ast\Delta\right)\ast\mathcal{T}_N=G_N$, where $\ast$ stands for the time convolution along the KB contour.
	
	\subsection{The adiabatic preparation of the initial state}
	\label{sec:adi}
	
	In this section, we describe how the initial state before the laser pulse can be prepared numerically efficiently within the present framework.
	First, we assume that the electrons are confined within the emitting tip at equilibrium. Thus, it is reasonable to regard the vacuum medium as empty at $t=0$, which implies $g^<\sim0$. Moreover, we
	can assume that the state reached by the system after applying the voltage difference $V_{pot}$ is adiabatically prepared from the equilibrium state, such that $J\left(x,t=0\right)=0$. This assumption is supported by the fact that an exponentially small steady-state current is expected when the tip is voltage-biased~\cite{fowler1928electron}.
	Under these assumptions, the imaginary-time components can be eliminated from the Dyson equation for the lesser component of $\mathbb{G}$ (see Appendix~\ref{app:B} for a detailed derivation):
	\begin{align}
		\mathbb{G}^{<}\left(X,X'\right)&=\int_{0}^{t_{max}}dt_{1}dt_{2}\int_{0}^{\infty}dx_{1}dx_{2}[g^{R}\left(X,X_{1}\right)\times\nonumber\\\times&J\left(X_{1}\right)\mathcal{T}_{N}^{<}\left(t_{1},t_{2}\right)J^{*}\left(X_{2}\right)g^{A}\left(X_{2},X'\right)],
	\end{align}
	where $t_{max}$ is the (real) time up to which we study the evolution of the system, and $g^R$ ($g^A$) stands for the retarded (advanced) component. Then, considering the emission process as exponentially decaying over space for $x\geq0$, we model the
	tunneling matrix element as $J\left(x,t\right)=J_{0}(t)\delta\left(x-0^{+}\right)$. In our chosen gauge, where the pulse-induced perturbation enters through the time-dependent potential energy $\mathcal{U}$, $J_0$ depends on time only through the adiabatic preparation of the initial state. Accordingly, we set $J_0\left(t\geq\tau_{ad}\right)=J_c$, where $\tau_{ad}\ll t_1$ denotes the characteristic timescale of the adiabatic preparation, and $J_c$ is the stationary single-particle current density established at the end of this process at the matter-vacuum interface. Under these assumptions, the lesser Green's function takes the form
	\begin{align}
		\mathbb{G}^{<}\left(X,X'\right)&=\int_{0}^{t_{max}} dt_{1}dt_{2}g^{R}\left(X,\chi_{1}\right)J_{0}\left(t_{1}\right)\times\nonumber\\\times&\mathcal{T}_{N}^{<}\left(t_{1},t_{2}\right)J_{0}^{*}\left(t_{2}\right)g^{A}\left(\chi_{2},X'\right)\label{eq:full_G_Jdelta},
	\end{align}
	with $\chi_{i}=\left(x_{i}=0,t_{i}\right)$.
	Consequently, we identify the far-field region with $x\geq\bar{x}$, where $\bar{x}>0$ is chosen sufficiently large that all field-induced effects become negligible for $x\geq\bar{x}$ (as discussed also in Subsec.~\ref{subsec:vac_barr_prof}).
	In this region, the electronic states are well approximated by plane waves and, thus, the electronic momentum $\mathbf{k}=k\hat{\mathbf{x}}$ is a good quantum number. Therefore, we Fourier-transform $\mathbb{G}^{<}\left(X,X'\right)$, given by Eq.~\eqref{eq:full_G_Jdelta}, with respect to the spatial coordinates for $x,x'\geq\bar{x}$:
	\begin{equation}
		\mathbb{G}_{k}^{<}\left(t,t'\right)\equiv\int_{\bar{x}}^{\infty}dx\,dx'\,e^{-ikx}\,e^{ikx'}\,\mathbb{G}^{<}\left(X,X'\right).\label{eq:G_k_FT_def}
	\end{equation}
	In doing so, we obtain
	\begin{equation}
		\mathbb{G}_{k}^{<}=\bar{g}_{k}^{R}\ast\bar{\mathcal{T}}_{N}^{<}\ast\bar{g}_{k}^{A},\label{eq:G_{k}}
	\end{equation}
	where
	\begin{equation}
		\bar{g}_{k}^{R}\left(t,t'\right)\equiv\int_{\bar{x}}^{\infty}dx\,e^{-ikx}\,g^{R}\left(X,\chi'\right),\label{eq:gbar_k_def}
	\end{equation}
	\begin{equation}
		\bar{\mathcal{T}}_{N}\left(t,t'\right)=J_{0}\left(t\right)\mathcal{T}_{N}\left(t,t'\right)J_{0}^{*}\left(t'\right),
	\end{equation}
	and $\bar{g}_{k}^{A}\left(t',t\right)=\left[\bar{g}_{k}^{R}\left(t,t'\right)\right]^{*}$, consistently with the general retarded-advanced relation~\cite{RevModPhys.86.779}.
	Finally, the time-integrated mean far-field electron energy distribution can be written as
	\begin{equation}
		\left\langle n\right\rangle \left(\varepsilon\right)=-i\sum_{k\,:\,\varepsilon_{ff}(k)=\varepsilon}\left[\lim_{t_{max}\rightarrow+\infty} \mathbb{G}_{k}^{<}\left(t_{max}, t_{max}\right)\right],\label{eq:n_eps}
	\end{equation}
	i.e., the sum runs over all $k$-values mapping onto the same far-field energy $\varepsilon=\varepsilon_{ff}(k)$. Therefore, it emerges that the far-field energy distribution can be interpreted directly in terms of the features of the emitter contained in $\mathcal{T}_{N}^{<}$, filtered in the far-field distribution through isolated vacuum propagators.
	
	\subsection{The modified ``zipper algorithm''}
	
	At this stage, we briefly outline the algorithm used to simulate the proposed model. All the above-mentioned processes can be summarized in a procedure similar to the ``zipper algorithm'' described in~\cite{Potthoff_PhysRevB.59.2549, PhysRevB.70.195342, PhysRevB.88.075135}:
	\begin{enumerate}
		\item Determine the needed time- and space-dependent retarded propagator for the isolated vacuum, that is, $g^R\left(X,\chi'\right)$ (see Appendix~\ref{app:A}).
		\item Given a single-electron
		tunneling matrix element $J_{0}=J_{0}\left(t\right)$, build the retarded hybridization function $\Delta^R=J_{0}\left(t\right)g^R\left(\chi,\chi'\right)J_{0}\left(t'\right)$.
		\item Solve the equation of motion for the diagonal elements $G_{i,i}$ along the matter chain:
		\begin{enumerate}
			\item Choose the left boundary condition ($\Delta_{0}^{l}$) and a guess for each diagonal element $G_{i,i}$.
			In this work, we implement the bulk boundary conditions for the metallic chain~\cite{PhysRevB.88.075135}, i.e., $\Delta_{0}^{l}=\Delta_{1}^{l}$.
			\item Compute the local self-energies $\Sigma_{i}$ for $i=1,\ldots,N$.
			\item Update the left hybridization functions from $1$ to $N$, since $\Delta_{i}^{l}$ depends on $\Delta_{i-1}^{l}$~\cite{PhysRevB.88.075135}:
			\begin{equation}
				\Delta_{i}^{l}=t_{i}^{*}\frac{1}{\mathcal{G}_{i}^{-1}-\Delta_{i-1}^{l}}t_{i}.
			\end{equation}
			\item Update the right hybridization functions from $N$ to $2$, since $\Delta_{i}^{r}$ depends on $\Delta_{i+1}^{r}$~\cite{PhysRevB.88.075135}:
			\begin{equation}
				\Delta_{i}^{r}=t_{i-1}\frac{1}{\mathcal{G}_{i}^{-1}-\Delta_{i+1}^{r}}t_{i-1}^*.
			\end{equation}
			
			The connection with the vacuum emerges at this stage via the term
			\begin{equation}
				\label{eqbcright}
				\Delta_{N+1}^r = \Delta.
			\end{equation}
			
			\item Update the diagonal elements $G_{i,i}$ for $i=1,\ldots,N$ via Eq.~\eqref{eq:dys_eq_trid}.
			\item Repeat the procedure from step (b) until convergence is reached for each $G_{i,i}$.
		\end{enumerate}
		\item Once $G_{N}$ is inferred, solve the integral equation
		$\left(1-G_{N}\ast\Delta\right)\ast\mathcal{T}_N=G_N$ (which is equivalent to Eq.~\eqref{aghsjssass}) for $\mathcal{T}_N$.
		\item Finally, compute $\mathbb{G}^<$ in the far-field region via Eq.~\eqref{eq:full_G_Jdelta} and, thus, the far-field energy distribution.
	\end{enumerate}
	Steps 2--4 of this algorithm require calculations along the $L$-shaped contour, such as solving KB equations, for which we use the NESSi library~\cite{SCHULER2020107484}.
	
	\subsection{The perturbative tunneling regime}\label{subsec::IID}
	
	The key result in Eq.~\eqref{eq:G_{k}} becomes more transparent in the
	perturbative tunneling regime, where the laser pulse and the tunneling
	process are assumed to have no ``back-action'' on the metallic tip, implying
	$\Delta^{R}=0$. It follows that Eq.~\eqref{aghsjssass} reduces to $\mathcal{T}_{N}=G_{N}$, and that the right boundary condition in the zipper algorithm for $G_N$, Eq.~\eqref{eqbcright}, is given by
	$\Delta^r_{N+1} =0$. Under this assumption, $G_{N}^{<}$ is a time-translational invariant quantity that can be Fourier-transformed in frequency space:
	\begin{equation}
		G_{N}^{<}\left(\omega\right)=\int dt{G}_{N}^{<}\left(t\right)e^{i\omega t}.
	\end{equation}
	Moreover, we can filter out the far-field free-electron propagators $\underline{g}$ from the full ones as follows: $\bar{g}_{k}^{R}\left(t,t'\right)=e^{-i\varepsilon_{ff}\left(k\right)\left[t-t'\right]}\underline{g}_{k}^{R}\left(t,t'\right)$.
	
	Then, we introduce, for $t'\rightarrow+\infty$, the single-time modified vacuum propagator
	\begin{equation}
		\label{eq:skt}
		s_{k}\left(t\right)=\lim_{t'\rightarrow+\infty}J_0\left(t\right)\underline{g}_{k}^{R}\left(t',t\right),
	\end{equation}
	and its Fourier amplitude
	\begin{equation}
		\label{eq:skw}
		s_{k}\left(\omega\right)=\int dt\, s_{k}\left(t\right)e^{i\omega t}.
	\end{equation}
	We further define the field-dressed vacuum spectral weight as
	\begin{equation}
		\label{eq:sk_spectral_weight}
		W_v\left(\omega,\varepsilon\right) = \sum_{k\,:\,\varepsilon_{ff}(k)=\varepsilon} \left|s_{k}\left(\omega\right)\right|^{2}.
	\end{equation}
	When inserting the last expressions into
	Eqs.~\eqref{eq:full_G_Jdelta}--\eqref{eq:G_{k}}, we obtain
	\begin{equation}
		\label{eq:n_eps_sk_omega}
		\left\langle n\right\rangle \left(\varepsilon\right)=\int d\omega\,
		W_v\left(\omega,\varepsilon\right)\,\mathcal{N}_{N}\left(\varepsilon-\omega\right),
	\end{equation}
	where
	$\mathcal{N}_{N}=\Im\left\{ G_{N}^<\right\}/2\pi$ is the density of occupied states on the last matter site $N$.
	
	From Eq.~\eqref{eq:skt}, we observe that, for $t\geq\tau_{ad}$, $s_{k}\left(t\right)$ depends on time only through the perturbation induced by the laser pulse. As a consequence, its Fourier amplitude $s_k\left(\omega\right)$ is expected to present peaks at the laser frequency $\omega_{ph}$ and its integer multiples, broadened by the finite pulse envelope: $s_{k}\left(\omega\right)\sim\sum_{n}a_{k,n}\,\delta\left(\omega-n\omega_{ph}\right)$. Therefore, we have that
	\begin{equation}
		\label{eq:sk_spectral_weight_appr}
		W_v\left(\omega,\varepsilon\right) \sim\sum_{n}\alpha_{n}\left(\varepsilon\right) \delta\left(\omega-n\omega_{ph}\right),
	\end{equation}
	with $\alpha_{n}\left(\varepsilon\right)=\sum_{k\,:\,\varepsilon_{ff}(k)=\varepsilon} \left|a_{k,n}\right|^{2}$.
	Following on from this consideration, we finally infer
	\begin{equation}
		\label{eq:n_vs_en_formula}
		\left\langle n\right\rangle \left(\varepsilon\right)\sim\sum_{n}\alpha_{n}\left(\varepsilon\right)\,\mathcal{N}_{N}\left(\varepsilon-n\omega_{ph}\right).
	\end{equation}
	
	It follows that, in a perturbative tunneling regime, the far-field energy distribution can be interpreted as the result of photon exchange between the impinging laser pulse and the electrons in the tip.
	The latter occupy, at equilibrium, energy levels consistent with the local density of occupied states (LDOS) $\mathcal{N}_{N}$.
	Hence, the equilibrium LDOS can be seen as transferred to the far-field energy distribution after being shifted by integer multiples of the photon energy. In addition, it is modulated by the amplitudes of propagators in frequency space, which, in turn, encode features of the barrier profile and the laser pulse.
	
	\section{Results\label{sec:Results}}
	
	\subsection{Vacuum barrier profile}
	\label{subsec:vac_barr_prof}
	
	A consistent description of the pulse-induced perturbation is required to
	infer the energy distribution of electrons emitted from a voltage-biased
	nanotip following excitation by an ultrafast pulse at the matter-vacuum
	interface. The impingement of the pulse is typically modeled by writing the
	electric field in the vacuum as the sum of an oscillating laser field
	$\boldsymbol{\mathcal{E}}$ and a static field $\boldsymbol{F}$ (barrier
	field) associated with the applied voltage difference
	$V_{pot}$~\cite{Zhang2016, PhysRevB.88.195434}. By contrast, the
	penetration of the laser into the emitting material is commonly neglected or
	treated as exponentially decaying~\cite{PhysRevLett.105.147601}. Therefore, by assuming that the pulse does not penetrate the tip ($x<0$), we choose the
	vacuum barrier profile $\mathcal{U}(x,t)$ as
	\begin{equation}
		\mathcal{U}\left(x\geq0,t\right)=\phi-V_{pot}\left[1-\exp\left(-\frac{\left(F+\mathcal{E}\left(x,t\right)\right)x}{V_{pot}}\right)\right],\label{eq:V_x_t_prof}
	\end{equation}
	where $\phi$ is the emitter work function, $\boldsymbol{F}=F\hat{\mathbf{x}}$,
	and $\boldsymbol{\mathcal{E}}(x,t)=\mathcal{E}(x,t)\hat{\mathbf{x}}$, with
	\begin{align}
		\mathcal{E}\left(x,t\right)&=\mathcal{E}_{0}\left[1+\left(\gamma_{0}-1\right)\exp\left(-\frac{x}{\lambda_{nf}}\right)\right]\times\nonumber\\
		\times&\exp\left(-\ln2\frac{x^{2}}{\lambda_{g}^{2}}\right)\exp\left(-4\ln2\frac{\left(t-t_{cen}\right)^{2}}{\tau_{p}^{2}}\right)\times\nonumber\\
		\times&\cos\left(\frac{2\pi}{T_{per}}\left(t-t_{cen}\right)\right)\label{eq:E_{x}{}_{t}}
	\end{align}
	describing a Gaussian laser beam focused on a small region at the edge of
	the nanotip~\cite{PhysRevLett.96.077401, PhysRevLett.105.147601,
		PhysRevLett.97.247402}. Its amplitude $\mathcal{E}_{0}\geq0$ is enhanced
	near the tip apex by a factor $\gamma_{0}>1$, which accounts for the
	field-enhancement effect~--~a technologically relevant phenomenon that
	occurs when the tip thickness is smaller than the laser
	wavelength~\cite{AVErmushev_1993, thomas2015large, Piltan_2018}. In addition,
	$\lambda_{nf}$ is the decay length associated with this
	effect~\cite{Seiffert_2018}, and $\lambda_{g}>\lambda_{nf}$ is the
	half-width at half-maximum of the focused laser spot. Finally, the
	field oscillates in time with angular frequency $\omega_{ph}=2\pi/T_{per}$,
	modulated by a Gaussian envelope of full-width at half-maximum $\tau_{p}$
	centered at $t_{cen}$.
	
\begin{figure}
		\centering
		\includegraphics[width=0.97\columnwidth]{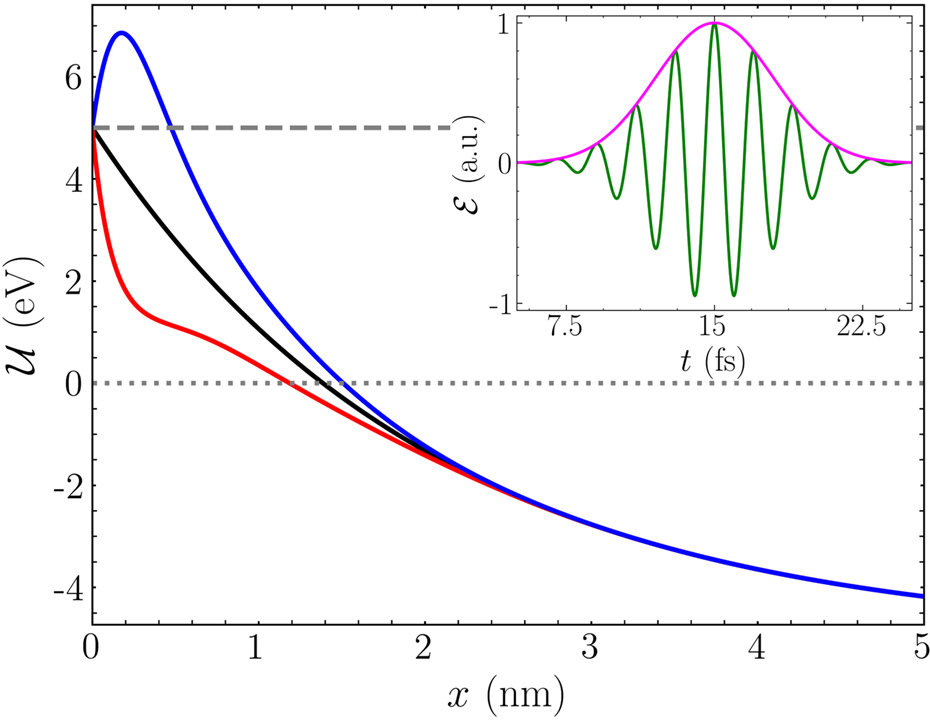}
		\caption{Vacuum barrier profile near the matter-vacuum interface at three different instants of time: equilibrium ($t = 0$, black), field maximum ($t = t_{cen}$, red), and field minimum ($t = t_{cen} + T_{per}/2$, blue). The following parameters are used: $V_{pot} = 10\,\,\mathrm{V}$, $F = 5\,\,\mathrm{V/nm}$, $\mathcal{E}_0 = 2\,\,\mathrm{V/nm}$, $\gamma_0 = 15$, $\lambda_{nf} = 0.25\,\,\mathrm{nm}$, $\lambda_g = 1\,\,\mathrm{nm}$, $t_{cen} = 15\,\,\mathrm{fs}$, $\tau_p = 7\,\,\mathrm{fs}$, and $T_{per} = 2\,\,\mathrm{fs}$. The Fermi level ($\mu = 0\,\,\mathrm{eV}$) and work function ($\phi = 5\,\,\mathrm{eV}$) are denoted with dotted and dashed gray lines, respectively. The inset shows the normalized temporal evolution of the pulse electric field (green) modulated by a Gaussian envelope (magenta), both in arbitrary units.}
		\label{fig:time-dep-vac-barr-prof}
	\end{figure}

	The vacuum barrier profile in Eq.~\eqref{eq:V_x_t_prof}, shown in
	Fig.~\ref{fig:time-dep-vac-barr-prof} for representative parameter values, encodes two physically relevant limits. First, close to the interface
	($0\leq x\ll1$), it reduces to the FN triangular-barrier form,
	\begin{equation}
		\mathcal{U}\left(0\leq x\ll1,t\right)\simeq\phi-\left(F+\mathcal{E}\left(x,t\right)\right)x,
	\end{equation}
	in which the laser field acts as an effective, time-dependent correction
	to the static barrier field. Conversely, far from the interface
	($x\gg1$), where the field-induced effects become negligible,
	Eq.~\eqref{eq:V_x_t_prof} recovers the flat asymptotic value
	$\mathcal{U}(x\gg1,t)\to\phi-V_{pot}$, consistent with the requirement that
	electrons propagate as free particles in the far-field region.
	
	Given the barrier profile, we determine the starting point $\bar{x}$ of
	the far-field region, where the energy distribution of the emitted
	electrons is measured. In this region, as previously noted, the
	single-electron energy dispersion follows the free-electron parabola
	shifted by $\phi - V_{pot}$. Then, for the boundary $\bar{x}$ of
	the far-field region (used for the projection onto the final-state energy
	distribution in Eq.~\eqref{eq:n_eps}), we require that the following two
	conditions hold simultaneously:
	\begin{enumerate}
		\item the incident laser pulse has no significant influence, i.e., $\frac{\left|\mathcal{E}\left(x,t\right)\right|}{F}<\rho_{thr}$;
		\item the bending of the vacuum barrier due to the applied voltage can be neglected, i.e., $\exp\left(-\frac{Fx}{V_{pot}}\right)<\rho_{thr}$.
	\end{enumerate}
	Here, $\rho_{thr}\ll1$ denotes a threshold value that must be chosen
	sufficiently small.
	
	\subsection{Vacuum discretization and adiabatic switch-on}
	
	After discretizing the space in the region $x>0$, the vacuum essentially becomes a one-dimensional, uncorrelated tight-binding chain with $M$ sites, with only nearest-neighbor hopping. The hopping amplitude between adjacent sites is denoted by $t_v$, and the on-site energy at site $i$ and time $t$ is given by $\epsilon_v\left(i,t\right)=\mathcal{U}\left(x=\left(i-1\right)b,t\right)+2t_v$. Here, $b=1/\sqrt{t_{v}}$ is the corresponding lattice constant. Within this framework, the far-field single-electron energy dispersion becomes
	\begin{equation}
		\varepsilon_{ff}\left(k\right)=2t_{v}\left[1-\cos\left(kb\right)\right]+\phi-V_{pot}.
	\end{equation}
	
\begin{figure}
		\centering
		\includegraphics[width=0.97\columnwidth]{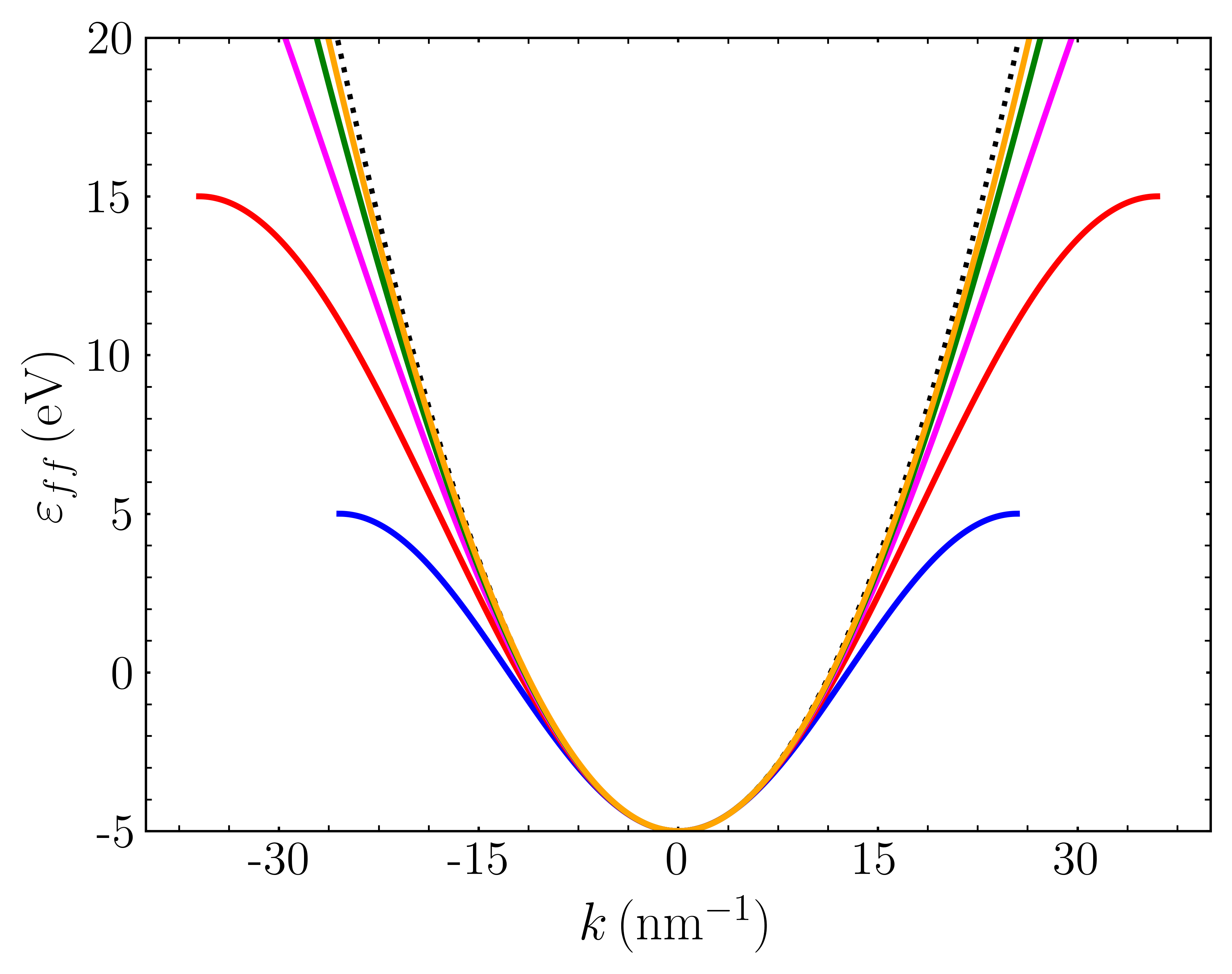}
		\caption{ Comparison of the far-field free-electron parabola (dotted black line) with tight-binding energy dispersions obtained for $t_{v}=2.5\,\,\mathrm{eV}$ (blue), $t_{v}=5.0\,\,\mathrm{eV}$ (red), $t_{v}=10\,\,\mathrm{eV}$ (magenta), $t_{v}=20\,\,\mathrm{eV}$ (green), and $t_{v}=40\,\,\mathrm{eV}$ (yellow). The discrepancy between the parabolic and tight-binding dispersions worsens at higher momenta and decreases as $t_{v}$ increases.}
		\label{fig:free_ele_par_vs_tb_disp}
	\end{figure}

	The accuracy of the tight-binding model in reproducing the continuum limit depends on the value of $t_v$: the higher its value, the wider the energy range over which the tight-binding energy dispersion provides a good approximation to the free-electron parabola (see Fig.~\ref{fig:free_ele_par_vs_tb_disp}).
	
	In addition, according to the proposed tight-binding picture, the
	tunneling matrix element $J_0=J_0\left(t\right)$ is mapped into a hopping amplitude $t_c=t_c\left(t\right)$ between the last site of the emitter and the first site in the vacuum.
	We choose
	\begin{equation}
		t_c\left(t\right)=\begin{cases}
			0 & \text{for}\,\,\,\,\,t\leq0\\
			t_c\sin^{4}\left(\frac{\pi t}{2\tau_{ad}}\right) & \text{for}\,\,\,\,\,0<t\le\tau_{ad}\\
			t_c & \text{for}\,\,\,\,\,t>\tau_{ad}
		\end{cases},
	\end{equation}
	where the second term is the adiabatic switch-on of matter-vacuum hopping for preparing the initial state (see Sec.~\ref{sec:adi}). In particular, it allows electrons to hop from the last matter site to the first vacuum site for $0\leq t \leq \tau_{ad}$, generating an unphysical electronic bunch that propagates in the vacuum. Later, when the laser pulse impinges at time $t_1$, additional electrons are emitted from the metallic tip, and their energy distribution is computed in the far-field region. To isolate the pulse-induced contribution, we perform two simulations for each case: one with the pulse and one without. If $\tau_{ad}$ is sufficiently smaller than $t_1$, we remove artifacts generated during $0\leq t \leq \tau_{ad}$ as follows:
	\begin{equation}
		\label{eq:subtr_pop}
		\left\langle n\right\rangle =\left\langle n\right\rangle ^{w.p.}-\left\langle n\right\rangle ^{n.p.},
	\end{equation}
	where $\left\langle n\right\rangle ^{w.p.}$ and $\left\langle n\right\rangle ^{n.p.}$ are the time-integrated far-field energy distributions with and without the pulse, respectively. This procedure ensures that only the physical features arising from the pulse remain.
	
	\subsection{The case without electronic correlations\label{subsec::IIC}}
	
	As a first benchmark for the NEGF framework, we present results for the non-equilibrium electron emission from a metallic uncorrelated tip ($U=0$), consistent with the single-particle framework adopted in previous studies~\cite{PhysRevLett.96.077401, PhysRevLett.105.147601, PhysRevLett.97.247402, PhysRevLett.105.257601, PhysRevLett.107.087601, Seiffert_2018}.
	
	The following parameters are used: $\phi=5.0\,\,\mathrm{eV}$, $\beta=10\,\,\mathrm{eV}^{-1}$, $V_{pot}=10\,\,\mathrm{V}$, $t_{v}=10\,\,\mathrm{eV}$ (and, thus, $b=0.062\,\,\mathrm{nm}$), $\epsilon_{m}=0.0\,\,\mathrm{eV}$, $t_{m}=t_{c}=1.0\,\,\mathrm{eV}$, $t_{max}=55\,\,\mathrm{fs}$, $\tau_{ad}=1.0\,\,\mathrm{fs}$. For the laser pulse, we set $\tau_{p}=7.0\,\,\mathrm{fs}$, $t_{cen}=15\,\,\mathrm{fs}$, $T_{per}=2.0\,\,\mathrm{fs}$, and, thus, $\omega_{ph}=2.1\,\,\mathrm{eV}$; then, $\gamma_{0}=15$, $\lambda_{nf}=0.25\,\,\mathrm{nm}$ and $\lambda_{g}=1.0\,\,\mathrm{nm}$. By doing so, we assume that the pulse effects are confined to the vicinity of the matter-vacuum interface (see Fig.~\ref{fig:time-dep-vac-barr-prof}). Furthermore, we choose $\rho_{thr}=10^{-3}$ as the threshold value for constructing the far-field region.
	Finally, since the matter-vacuum hopping amplitude is significantly smaller than both the work function and the bandwidth of the matter chain, we restrict our simulation and analysis to the perturbative tunneling regime outlined in Subsec.~\ref{subsec::IID}.
	
	We begin by analyzing the time evolution of the electronic population along the vacuum chain, evaluated as $\langle n \rangle(x=(i-1)b,t) = -i\,\mathbb{G}^<_{i, i}\left(t,t\right)$ for $i = 1, \ldots, M$.
	
\begin{figure}
		\centering
		\includegraphics[width=0.97\columnwidth]{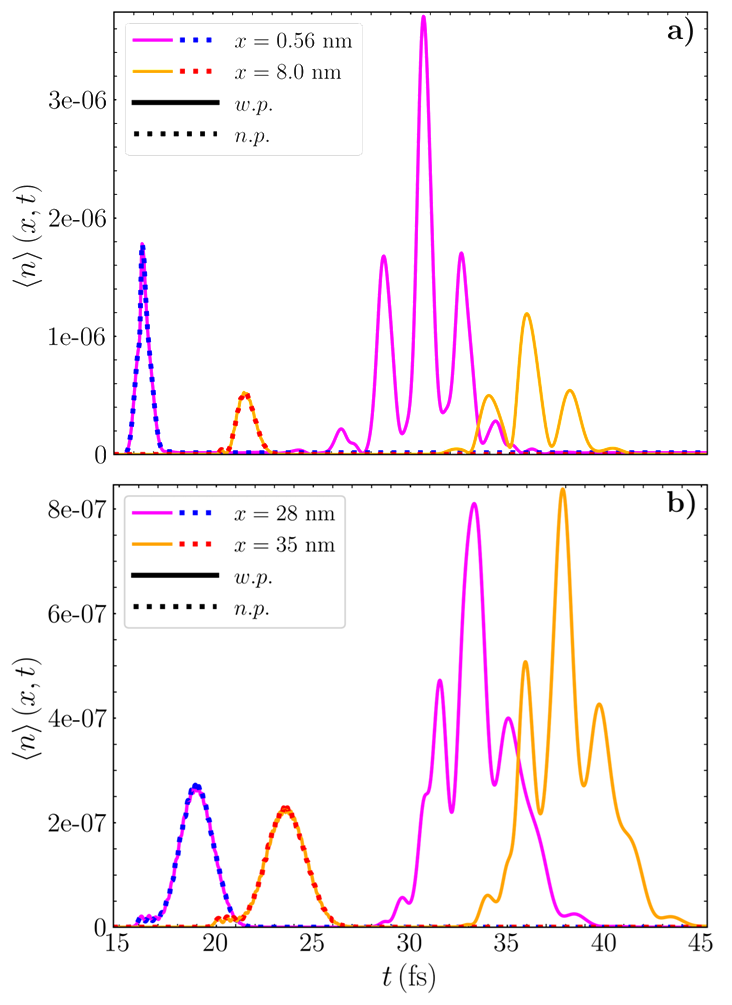}
		\caption{
			Time evolution of electronic populations close to the matter-vacuum interface \textbf{(a)} and in the far-field region \textbf{(b)}, shown both with (solid lines) and without (dotted lines) the laser pulse. Specifically, with the vacuum lattice constant set to $b=0.062\,\,\mathrm{nm}$, the following distances from the interface are considered: $x=0.56\,\,\mathrm{nm}$ (blue and magenta) and $x=8.0\,\,\mathrm{nm}$ (red and orange) in \textbf{(a)}; $x=\bar{x}=28\,\,\mathrm{nm}$ (blue and magenta) and $x=35\,\,\mathrm{nm}$ (red and orange) in \textbf{(b)}. In both cases, the artificial electronic bunches in the with- and without-pulse curves overlap in time and are therefore removed via the subtraction $\left\langle n\right\rangle(x,t) =\left\langle n\right\rangle^{w.p.}(x,t)-\left\langle n\right\rangle^{n.p.}(x,t)$.
		}
		\label{fig:pop_vs_t}
	\end{figure}

	The barrier field is fixed at $F = 5.0\,\,\mathrm{V}/\mathrm{nm}$.
	In the absence of a pulse, an artificial electronic bunch emerges at the interface ($\langle n \rangle^{n.p.}$). It propagates into the vacuum due to the adiabatic switch-on of the matter-vacuum coupling term (see Fig.~\ref{fig:pop_vs_t}). A laser pulse with electric field magnitude $\mathcal{E}_0 = 2.0\,\,\mathrm{V}/\mathrm{nm}$ is then applied.
	
	In this case, as shown in the same figure, the artificial bunch and the laser-induced emissions are temporally well-separated ($\langle n \rangle^{w.p.}$). Therefore, subtracting these quantities as reported in Eq.~\eqref{eq:subtr_pop} removes the unphysical background, isolating the contributions from the pulse both at the interface and in the far-field region (set for $x\geq\bar{x}=28\,\,\mathrm{nm}$). All results reported hereafter include this correction.
	
\begin{figure}
		\centering
		\includegraphics[width=0.97\columnwidth]{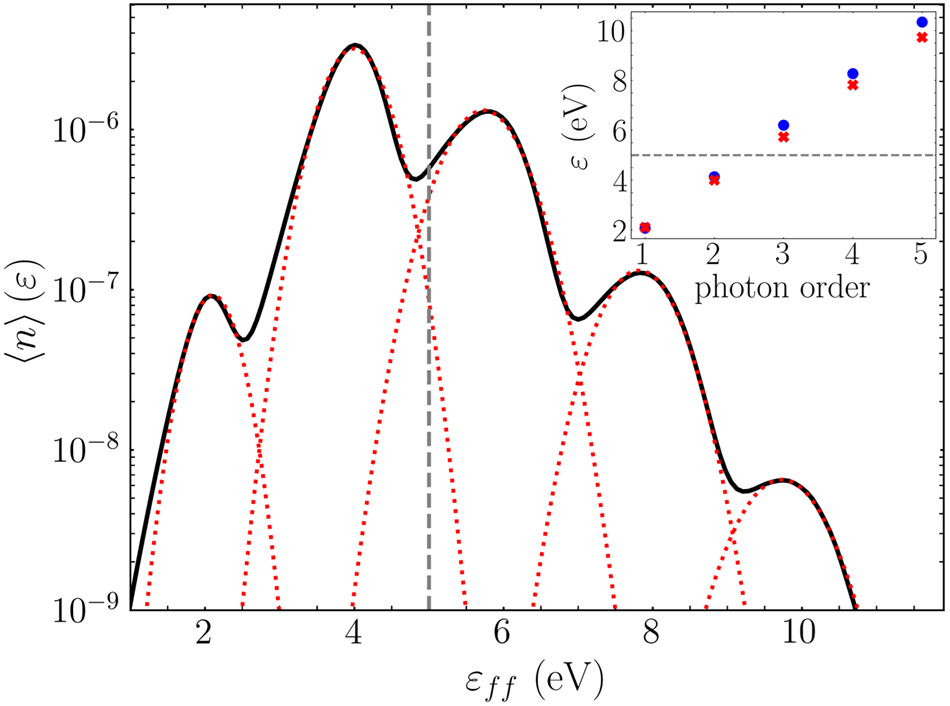}
		\caption{Time-integrated far-field electron energy distribution (black solid line), fitted with Gaussian functions (red dotted lines) centered at the arising peaks. In the inset, the peak energy positions obtained from the Gaussian fitting procedure (red crosses) are compared with integer multiples of the photon energy (blue circles), corresponding to different photon orders. In both panels, the work function ($\phi=5\,\,\mathrm{eV}$) is indicated by a dashed gray line.}
		\label{fig:ff_en_distr_and_ph_ord}
	\end{figure}

	Having validated our numerical procedure, we now focus on the time-integrated far-field electron energy distribution, depicted in Fig.~\ref{fig:ff_en_distr_and_ph_ord} on a logarithmic scale. Distinct peaks appear at specific energies, which we fit using Gaussian functions to extract their positions. The first five peaks align well with integer multiples of the photon energy, indicating that the observed structure results from absorption by electrons emitted from the tip. Stated differently, electrons within the tip absorb one or more photons from the pulse, gaining sufficient energy to either tunnel through the barrier (first two peaks) or overcome it (higher-order peaks). While the positions of the first two peaks correspond to $\varepsilon_{ff}=\mu+n\omega_{ph}$ for $n=1,2$, with the Fermi level $\mu=0$, those for $n=3,4,5$ are systematically shifted downward by approximately $0.5\,\,\mathrm{eV}$.
	
\begin{figure}
		\centering
		\includegraphics[width=0.97\columnwidth]{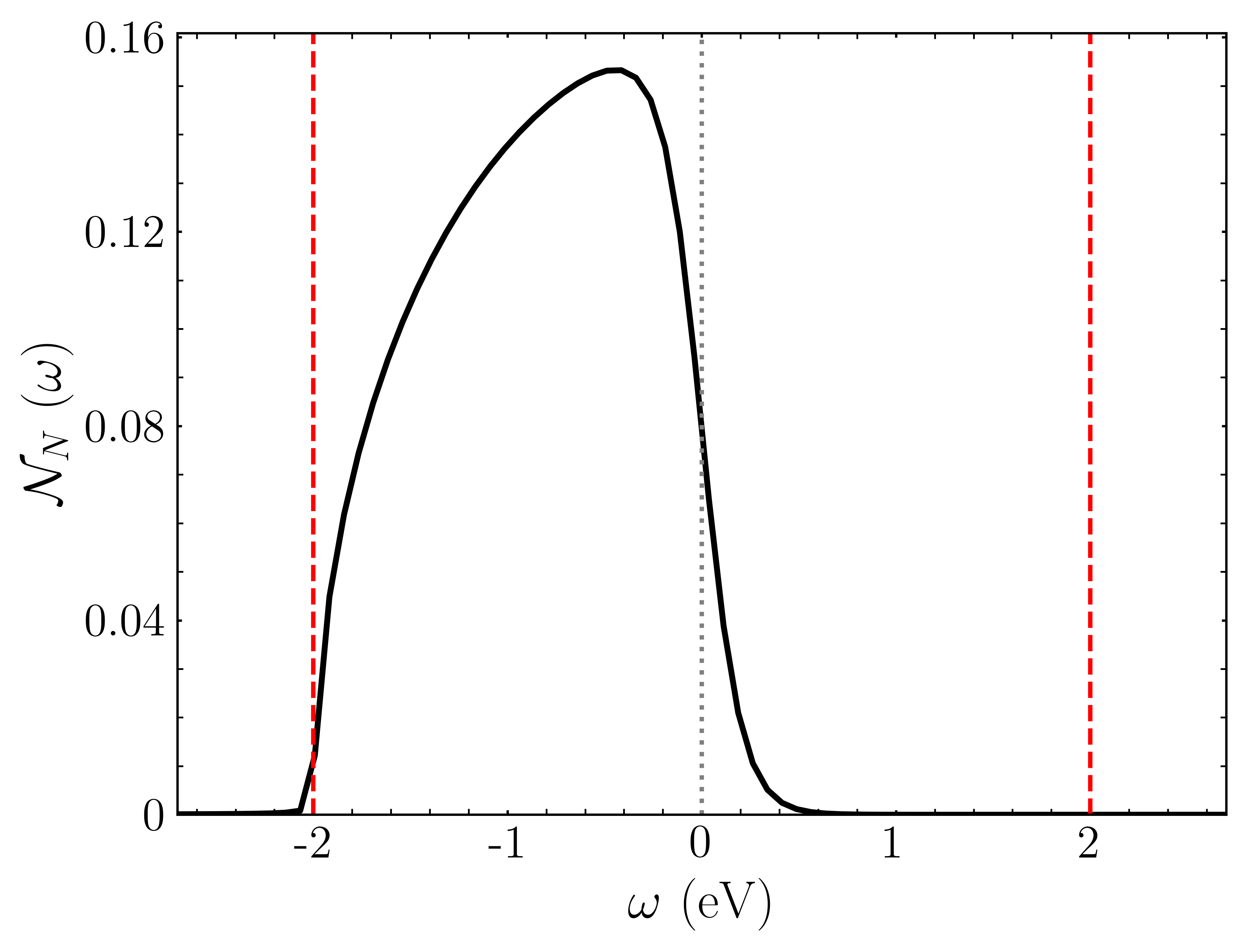}
		\caption{Density of occupied states on the last matter site ($\mathcal{N}_{N}$)
			plotted vs. energy in the uncorrelated case. The vertical dashed red lines mark the boundaries of the matter-energy dispersion
			($-2\,\,\mathrm{eV}$ and $2\,\,\mathrm{eV}$) when $U=0$, while the dotted gray one coincides with the Fermi level. \protect\label{fig:LDOS_unc}
		}
	\end{figure}

	This shift, consistent with previous findings~\cite{PhysRevLett.105.257601,
		PhysRevB.88.195434}, can be traced back to the structure of the local
	density of occupied states $\mathcal{N}_N$ itself (see
	Fig.~\ref{fig:LDOS_unc}), which is not uniformly distributed below the
	Fermi level, but instead has a peak approximately $0.5\,\,\mathrm{eV}$ below it.
	For $n=1,2$, the acquired energy is insufficient to overcome the barrier,
	so that emission proceeds via tunneling, whose probability favors
	electrons closer to the Fermi level ($\mu=0$); as a result, the spectrum
	in this regime is dominated by photofield emission (PFE)~\cite{Kruger2012,
		Zhou2019}. In contrast,
	higher-order processes ($n\geq3$) provide enough energy to overcome the
	barrier directly, allowing electrons from across the occupied continuum to
	contribute to multi-photon (MPP) or above-threshold photoemission
	(ATP)~\cite{Kruger2012, Zhou2019}, consistently with the underlying LDOS profile noted above.
	
	\subsection{Emission regimes within the perturbative framework}
	
	The physical interpretation presented in the previous subsection can be further refined by systematically varying either the pulse electric field amplitude $\mathcal{E}_{0}$ or the barrier field $F$ in the uncorrelated limit ($U=0$), while keeping all other parameters fixed. This analysis offers deeper insight into the impact of these two experimentally controllable quantities on the far-field energy distribution. By treating the matter-vacuum coupling perturbatively, the far-field energy distribution is determined by the local density of occupied states $\mathcal{N}_N$ and the field-dressed vacuum spectral weight $W_v$ (see Eq.~\eqref{eq:n_eps_sk_omega}).
	
\begin{figure}[h!]
		\centering
		\includegraphics[width=0.97\columnwidth]{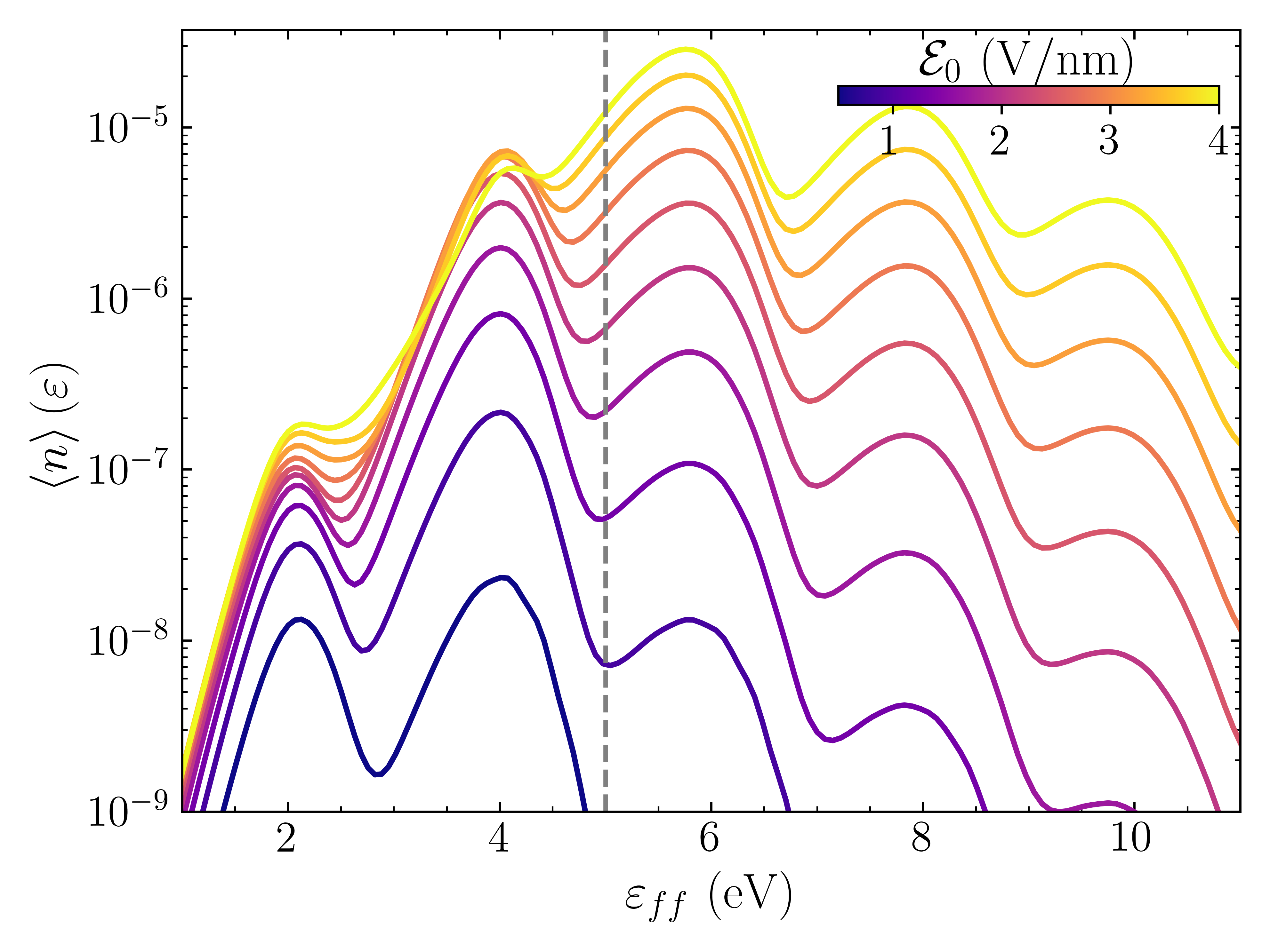}
		\caption{Far-field electron energy distributions for increasing pulse electric field amplitudes, ranging from $\mathcal{E}_0 = 0.50\,\,\mathrm{V/nm}$ to $\mathcal{E}_0 = 4.0\,\,\mathrm{V/nm}$. The static barrier field is fixed at $F = 5.0\,\,\mathrm{V/nm}$. As the pulse intensity increases, the probability of multi-photon excitations correspondingly rises, leading to the emergence of higher-order spectral features.}
		\label{fig:varying_E_0}
	\end{figure}

	In Fig.~\ref{fig:varying_E_0}, the time-integrated far-field energy distribution is shown for pulse electric field amplitudes $\mathcal{E}_{0}$ ranging from $0.5\,\,\mathrm{V/nm}$ to $4\,\,\mathrm{V/nm}$, with the barrier field fixed at $F=5\,\,\mathrm{V/nm}$. The energy positions of the peaks remain essentially unaffected by $\mathcal{E}_{0}$, confirming that they arise from photon absorption via the previously discussed mechanisms. In contrast, the height of the peaks strongly depends on $\mathcal{E}_{0}$. At low intensity ($\mathcal{E}_{0}=0.5\,\,\mathrm{V/nm}$), the emission is dominated by the photofield regime, with comparable one- and two-photon peaks. For intermediate intensities ($1\,\,\mathrm{V/nm} < \mathcal{E}_{0} < 2.5\,\,\mathrm{V/nm}$), the one-photon contribution diminishes in favor of two-photon processes, and the onset of multi-photon photoemission becomes evident. When $\mathcal{E}_{0}>3\,\,\mathrm{V/nm}$, high-order multi-photon and above-threshold photoemission channels dominate over the photofield emission, in agreement with previous results~\cite{PhysRevLett.97.247402}. This occurrence confirms that high-order photon absorption becomes significant only at sufficiently intense fields, allowing electrons to gain enough energy to overcome the barrier without tunneling.
	
	\begin{figure*}
		\centering
		\includegraphics[width=0.97\textwidth]{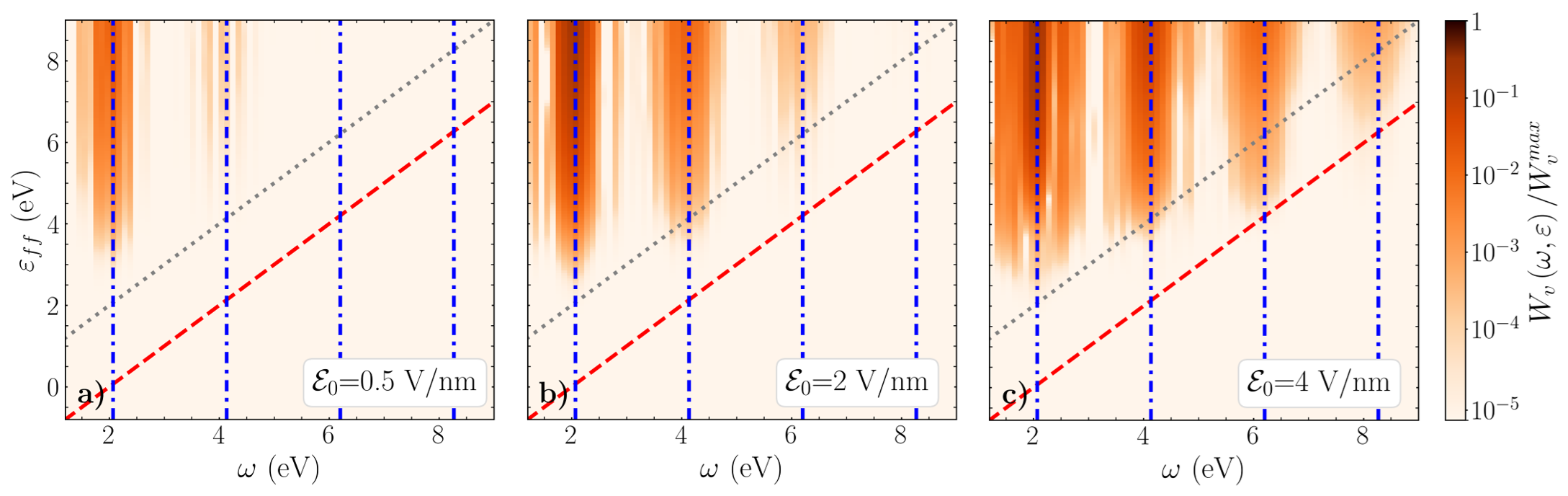}
		\caption{Field-dressed vacuum spectral weight $W_v$ as a function of $(\omega,\varepsilon_{ff})$ for
			$\mathcal{E}_{0}=0.50\,\,\mathrm{V/nm}$ \textbf{(a)}, $2.0\,\,\mathrm{V/nm}$
			\textbf{(b)}, and $4.0\,\,\mathrm{V/nm}$ \textbf{(c)}, with the barrier
			field fixed at $F=5\,\,\mathrm{V/nm}$. The amplitudes are normalized to the maximum value across the three field amplitudes. Four dash-dotted vertical blue lines are drawn in correspondence with $n\omega_{ph}$ for $n=1,2,3,4$, respectively.
			The dotted gray and dashed red lines
			denote the Fermi level and the lower edge of the matter-energy dispersion
			(see Fig.~\ref{fig:LDOS_unc}), respectively, each shifted upward by
			$\omega$: the region enclosed between them is where
			$\mathcal{N}_N(\varepsilon_{ff}-\omega)$ is nonzero and therefore
			contributes to $\langle n\rangle(\varepsilon)$.
		}
		\label{fig:vac_prop_E0}
	\end{figure*}
	
	Further insight is provided in Fig.~\ref{fig:vac_prop_E0}, which displays
	the field-dressed vacuum spectral weight $W_v$, defined in Eq.~\eqref{eq:sk_spectral_weight}, as a function of single-electron far-field energy, $\varepsilon_{ff}$, and frequency, $\omega$, for selected values of $\mathcal{E}_{0}$.
	Before Fourier-transforming, $s_k(t)$ is
	multiplied by a half-Hann window to suppress spectral leakage caused by truncation of the time integration. Moreover, as for the electronic populations (see Eq.~\eqref{eq:subtr_pop}), the field-dressed vacuum spectral weight also requires subtracting the no-pulse contribution, $W_v=W_v^{w.p.}-W_v^{n.p.}$. As expected, the spectra exhibit peaks at integer multiples of $\omega_{ph}$, confirming the validity of the analytical framework based on Eq.~\eqref{eq:n_vs_en_formula}.
	As $\mathcal{E}_{0}$ increases, high-order harmonics emerge, and their intensities scale proportionally with the field strength, thereby explaining the appearance of peaks in $\langle n \rangle(\varepsilon)$ above the work function for strong pulses. Notably, the first-order harmonic in $W_v \left(\omega,\varepsilon \right)$ remains the most prominent across all cases. However, the corresponding one-photon peak in the far-field energy spectrum is consistently weak. This apparent discrepancy can be explained by considering the role of the LDOS
	in Eq.~\eqref{eq:n_eps_sk_omega}. As reported in Fig.~\ref{fig:LDOS_unc}, it is nonvanishing only between $\sim-2\,\,\mathrm{eV}$ and the Fermi level and, thus, one-photon absorption promotes electrons to energy levels between $\sim0\,\,\mathrm{eV}$ and $\sim2\,\,\mathrm{eV}$. Yet, as shown in Fig.~\ref{fig:vac_prop_E0}, $W_v \left(\omega,\varepsilon \right)$ is suppressed in this energy range, explaining the observed restriction in one-photon emission even at high $\mathcal{E}_{0}$ values.
	
\begin{figure}
		\centering
		\includegraphics[width=0.97\columnwidth]{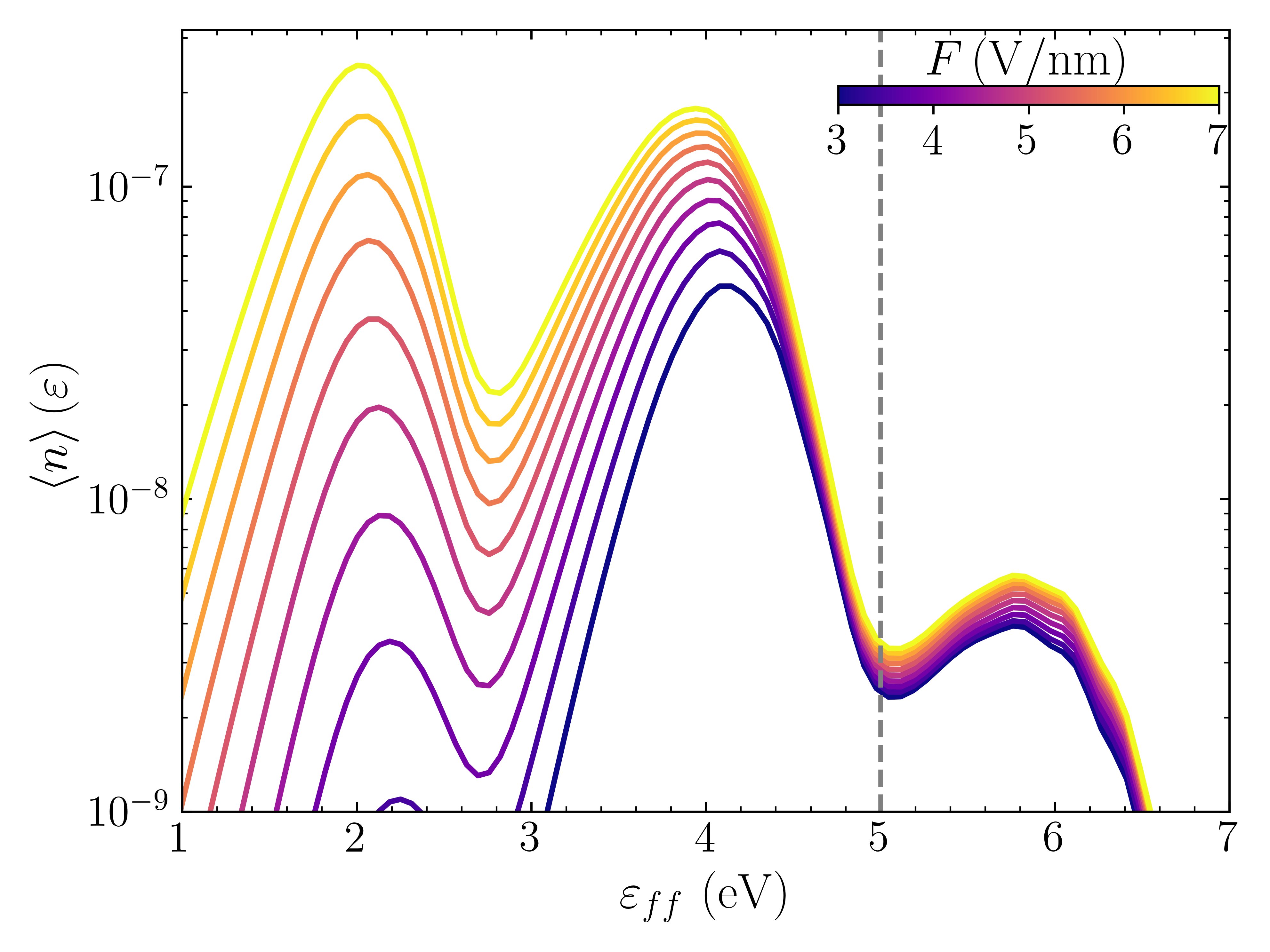}
		\caption{Far-field electron energy distributions obtained by varying the barrier field strength from $F = 3\,\,\mathrm{V/nm}$ to $F = 7\,\,\mathrm{V/nm}$, while keeping the pulse electric field fixed at $\mathcal{E}_0 = 0.75\,\,\mathrm{V/nm}$. Increasing the barrier field effectively narrows the tunneling barrier, enhancing the probability of emission processes dictated by single-photon absorption.
		}
		\label{fig:varying_F}
	\end{figure}

	The saturation of one- and two-photon processes can be better appreciated by analyzing how the far-field electron energy distribution varies with the barrier field $F$. Fig.~\ref{fig:varying_F} shows the far-field energy distributions for different barrier fields, equally spaced from $F=3\,\text{V/nm}$ to $F=7\,\text{V/nm}$, with a low pulse electric field amplitude $\mathcal{E}_0 = 0.75\,\text{V/nm}$ to minimize multi-photon contributions. The peaks above the work function $\phi$ exhibit minimal sensitivity to changes in $F$, supporting their interpretation as due to electron promotion over the barrier, which primarily depends on the barrier height and $\mathcal{E}_0$. In contrast, the one- and two-photon peaks are strongly influenced by $F$: as the barrier field increases, their amplitudes increase, and for $F > 6\,\text{V/nm}$, single-photon absorption dominates. This trend, also reported in~\cite{PhysRevB.81.115429}, supports the view that increasing $F$ steepens the barrier near the interface, thereby narrowing its width and facilitating photofield emission, especially when $\mathcal{E}_0$ is too low for triggering higher-order processes.
	
	\begin{figure*}
		\centering
		\includegraphics[width=0.97\textwidth]{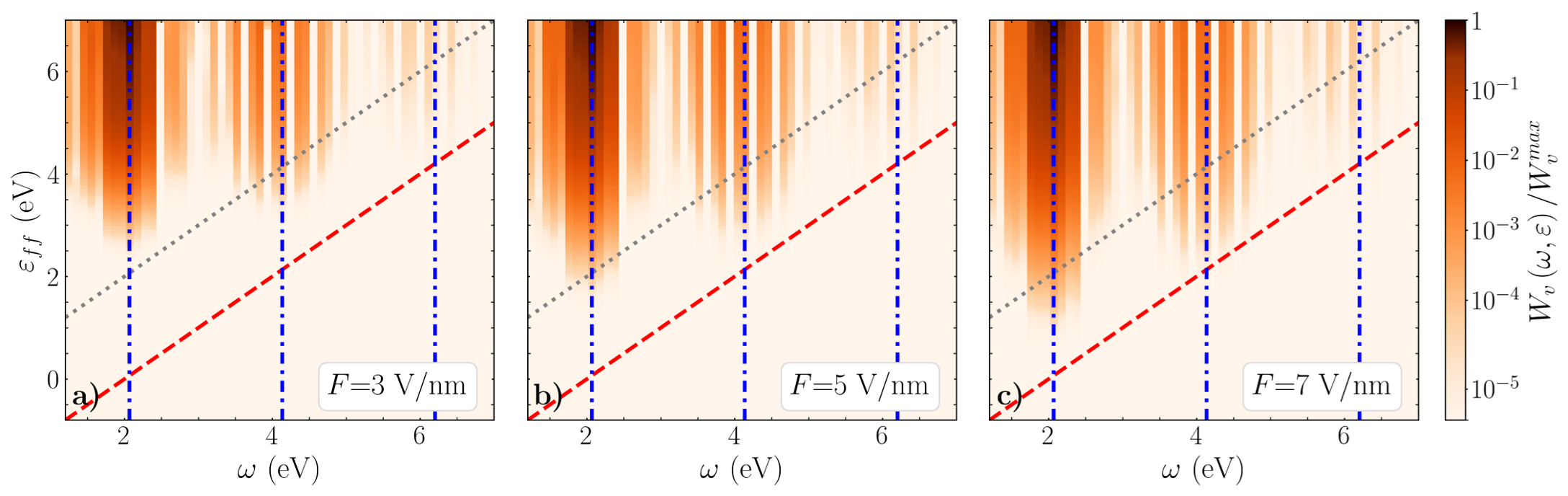}
		\caption{Intensity maps of $W_v(\omega,\varepsilon_{ff})$ for
			$F=3\,\,\mathrm{V/nm}$ \textbf{(a)}, $5\,\,\mathrm{V/nm}$ \textbf{(b)}, and
			$7\,\,\mathrm{V/nm}$ \textbf{(c)}, with the pulse amplitude fixed at
			$\mathcal{E}_{0}=0.75\,\,\mathrm{V/nm}$. The amplitudes are normalized to
			the global maximum across the three panels. Dash-dotted vertical blue lines mark $\omega=n\omega_{ph}$ for $n=1,2,3$, respectively. The dashed red and dotted gray lines
			have the same meaning as in Fig.~\ref{fig:vac_prop_E0}.}
		\label{fig:vac_prop_F}
	\end{figure*}
	
	\begin{figure}
		\centering
		\includegraphics[width=0.97\columnwidth]{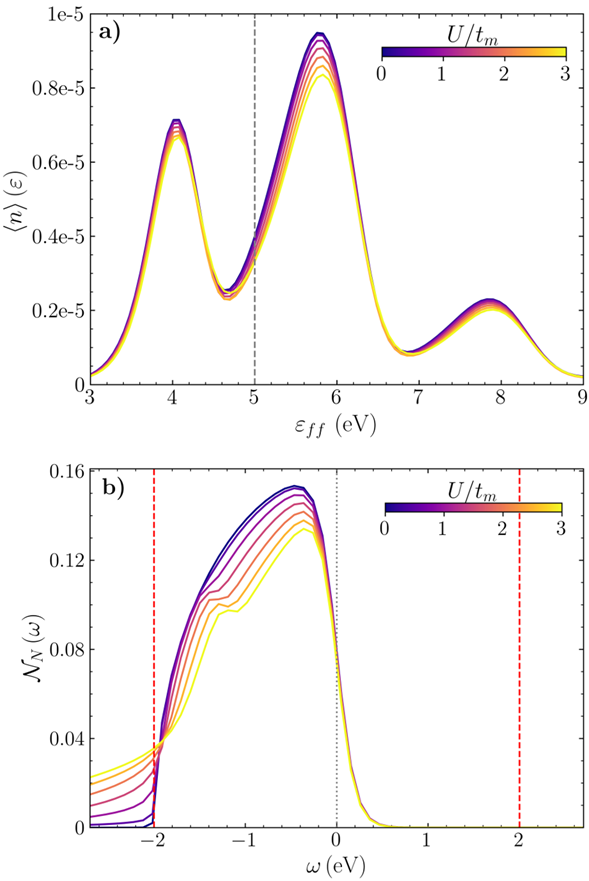}
		\caption{\textbf{(a)} One-, two-, and three-photon peaks in the far-field electron energy distribution as the ratio $U/t_m$ is varied from 0 (uncorrelated case) to 3, with fixed parameters $\mathcal{E}_0 = 3.0\,\,\mathrm{V/nm}$ and $F = 5.0\,\,\mathrm{V/nm}$. A vertical dashed gray line marks the work function. \textbf{(b)} Corresponding local densities of occupied states for each $U/t_m$ value. As the ratio increases, the LDOS decreases near the Fermi level (dotted gray line) and broadens in energy, extending beyond the limits of the uncorrelated matter energy dispersion (vertical dashed red lines).}
		\label{fig:varying_U_Nw}
	\end{figure}
	
	A more detailed understanding of this trend is provided, here as well, by examining the field-dressed vacuum spectral weight $W_v\left(\omega,\varepsilon\right)$, displayed in Fig.~\ref{fig:vac_prop_F} for $F=3\,\text{V/nm}$, $5\,\text{V/nm}$, and $7\,\text{V/nm}$. Let us focus on the energy range corresponding to one-, two-, and three-photon absorption, i.e., $0\,\,\mathrm{eV}\lesssim\varepsilon_{ff}\lesssim6\,\,\mathrm{eV}$.
	We observe that, when varying $F$, the amplitude of $W_v$ remains essentially unchanged for $\varepsilon_{ff}\gtrsim4\,\text{eV}$, where it attains its maximum along $\omega=\omega_{ph}$. Conversely, for $\varepsilon_{ff}\lesssim4\,\text{eV}$, $W_v$ grows markedly with $F$ along both $\omega=\omega_{ph}$ and $\omega=2\omega_{ph}$, progressively ``stretching'' downward in energy. This behavior underlies the growth of the one- and two-photon peaks with $F$ observed in Fig.~\ref{fig:varying_F}: at $F=7\,\text{V/nm}$, in particular, final states with $1\,\text{eV}\lesssim\varepsilon_{ff}\lesssim2\,\text{eV}$ become significantly populated at $\omega=\omega_{ph}$, signaling the dominance of single-photon absorption in this regime. Finally, the weaker higher-order harmonics exhibit no significant variation within this energy range as $F$ increases from $3\,\text{V/nm}$ to $7\,\text{V/nm}$, shown for $\omega=3\omega_{ph}$ in Fig.~\ref{fig:vac_prop_F}. This is consistent with the saturation of the three-photon peak in Fig.~\ref{fig:varying_F}, confirming that the barrier field, unlike the pulse intensity, does not affect the strengths of the MPP or ATP processes.
	
	\subsection{The weakly correlated case}
	
	In the previous subsections, we have shown how our numerical-theoretical framework efficiently captures laser-induced electron emission from a tip. We have identified distinct emission regimes and analyzed their dependence on parameters such as the barrier field and pulse intensity. Beyond offering a unified and versatile description of these phenomena, our approach explicitly retains the tip's role in the emission process. While the influence of the matter LDOS has already been discussed, its role becomes even more significant when electronic correlations within the tip are included, i.e., for $ U\neq 0$.
	
	In this work, since we deal with a metallic tip, we employ the self-consistent second-Born (2B) approximation, retaining the first- and second-order terms in $U$ for the local self-energy~\cite{Eckstein2018}:
	\begin{equation}
		\begin{cases}
			{\Sigma_{i}^{(1)}(t)=U\langle n_{i}\rangle(t)}\\
			{\Sigma_{i}^{(2)}(t,t')=U^{2}G_{i,i}(t,t')G_{i,i}(t,t')G_{i,i}(t',t)}
		\end{cases},
	\end{equation}
	where $i$ runs over all matter-site indices ($1, \ldots, N$). Even though we focus on a weak-correlation regime, we anticipate that electron-electron interactions will modify the matter LDOS. In turn, according to our framework, these correlation-induced modifications are expected to be transferred to the far-field electron energy distribution, leaving a mark of the electronic interactions in an experimentally observable quantity. In the following, we implement the same parameters and conditions as in Subsec.~\ref{subsec::IIC}, while setting $\mathcal{E}_{0}=3.0\,\,\mathrm{V/nm}$ and varying the ratio $U/t_{m}$ between 0 and 3.
	
	In Fig.~\ref{fig:varying_U_Nw}(a), we report the three main peaks of the far-field energy distribution for different values of $U/t_{m}$. Despite the weak interaction framework, electronic correlations manifest in the far-field energy spectra. As $U/t_{m}$ increases, we observe a global downward shift of the curves and a broadening of the valleys. These features can be traced back to modifications in the LDOS at equilibrium induced by interactions. Indeed, since the vacuum propagators are unaffected by $U$ and the pulse does not act on the matter in the perturbative tunneling regime, the observed changes must arise from the correlated LDOS.
	
	Fig.~\ref{fig:varying_U_Nw}(b) shows the LDOS for the same range of $U/t_{m}$: as the interaction increases, the LDOS near the Fermi level is suppressed and broadened in energy, with spectral weight extending beyond the non-interacting bandwidth. These correlation-induced modifications are transferred to the far-field energy distribution via the vacuum propagators, which, as discussed earlier, determine the relative strengths of the $n$-photon peaks and select the relevant portions of the LDOS that contribute to emission. As a result, electrons appear to be emitted from the weakly correlated metallic tip as if they were ``dressed'' by interactions. This interaction-induced dressing leaves a trace in their energy distribution, which remains detectable even far from the emission region.
	
	\subsection{The effects of the pulse's ``back-action'' on the emitter}
	
	So far, the laser-stimulated tunneling process has been treated within a
	perturbative framework. Under such an assumption, the extraction of
	electrons and the presence of an oscillating pulse have a negligible impact
	on the electronic structure of the metallic tip, so that the contour-ordered
	Green's function on the boundary site $N$ reduces to that of the isolated
	tip at equilibrium ($\mathcal{T}_N \simeq G_N$). Although this approximation
	is often sufficient for weak radiation-matter couplings, a more general
	treatment of strongly perturbed open quantum systems requires addressing the
	influence of field-driven effects back onto the emitter, since these can
	alter the electronic structure of the emitting material and thereby change
	its emission characteristics. For example, laser irradiation can induce a
	localized ultrafast heating of the electron gas~\cite{PhysRevB.88.195434},
	modifying the statistical Fermi-Dirac distribution and invalidating the perturbative tunneling description. To capture this pulse-induced
	``back-action'', we go beyond that approach by solving self-consistently the
	Dyson equation at the matter-vacuum interface,
	$(1 - G_N \ast \Delta) \ast \mathcal{T}_N = G_N$, i.e., by retaining the
	full right boundary condition $\Delta^{r}_{N+1} = \Delta$ of
	Eq.~\eqref{eqbcright} in the zipper algorithm.
	
\begin{figure*}
		\centering
		\includegraphics[width=1\textwidth]{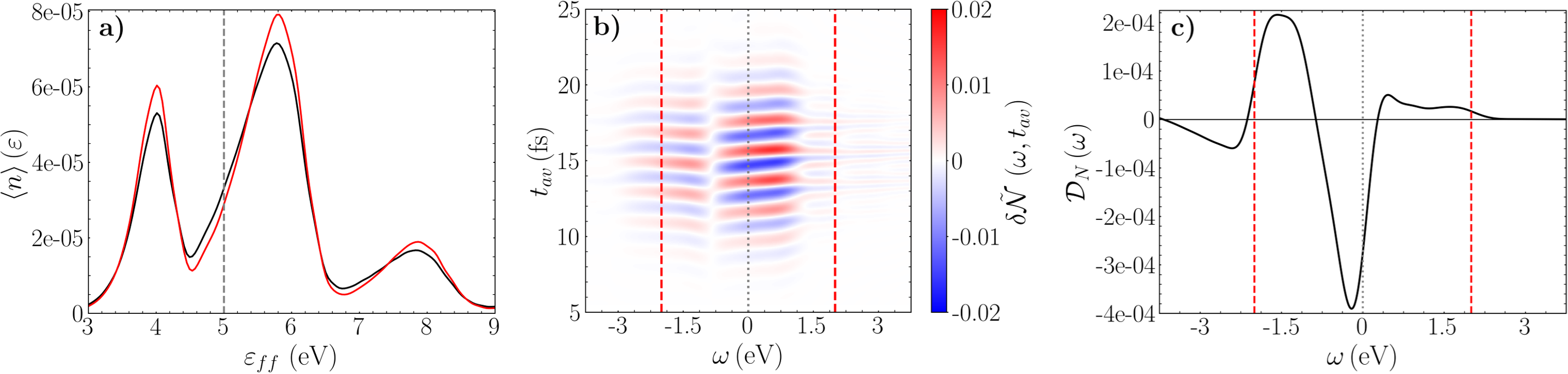}
		\caption{\textbf{(a)} Far-field electron energy distributions evaluated
			within the perturbative framework (red line) and by including the
			back-action of the pulse on the emitter (black line), for $U/t_{m}=1.5$,
			$t_{c}/t_{m}=3$, $\mathcal{E}_{0}=3$~V/nm, $F=5$~V/nm, and
			$\beta=30$~eV$^{-1}$. The dashed gray line marks the work function.
			Back-action induces peak broadening and partial filling of the valleys.
			\textbf{(b)} Pulse-induced variation of
			the local density of occupied states at the boundary site $N$,
			$\delta\tilde{\mathcal{N}}_{N}(t_{av},\omega)$. Red (blue) regions denote an
			increase (decrease) in the occupation of the levels at energy $\omega$. The
			lobes alternate with a period close to the optical one ($T_{per}=2$ fs). The dotted gray line marks the Fermi level and the dashed red
			lines the edges of the non-interacting matter dispersion. \textbf{(c)} Time-averaged variation
			$\mathcal{D}_{N}(\omega)$,
			Eq.~\eqref{eq:delta_bar}, obtained from the map in (b). Its structures
			account for the peak reduction and the valley
			filling observed in (a).}
		\label{fig:fig11}
	\end{figure*}

	Figure~\ref{fig:fig11}(a) compares the far-field emission profiles obtained
	with (black line) and without (red line) back-action, for $U/t_m=1.5$,
	$\mathcal{E}_0=3$~V/nm, and $F=5$~V/nm. The matter-vacuum coupling is set
	to $t_c/t_m=3$, a value larger than in Sec.~\ref{subsec::IIC}, chosen to
	drive the system beyond the perturbative regime, where the
	back-action effects become appreciable. In addition, to reduce the intrinsic thermal effects, the system is
	initialized at a lower temperature ($\beta = 30\,\text{eV}^{-1}$) than in the previous simulations. In the
	absence of back-action, the spectrum exhibits sharp multi-photon
	photoemission peaks separated by deep valleys. When the back-action is
	included, the peaks are systematically reduced in height, a continuous electron background partially fills the valleys, and the photonic
	sidebands are broadened toward lower energies. These modifications must
	originate from the hybridization function $\Delta\left(t,t'\right)$, which
	is the only term distinguishing the two calculations and which acts as an
	additional, time-dependent self-energy on the boundary site.
	
	To evaluate the dynamical effects mediated by $\Delta$, we consider the variation of the LDOS at the boundary site between
	the pulse-driven and the pulse-free evolution. Dealing with a non-equilibrium system, we compute the Wigner transform of
	the lesser Green's function,
	$\mathcal{T}_{N}^{<}(t,t') \rightarrow \tilde{\mathcal{T}}_{N}^{<}(t_{av},\omega)$,
	i.e., the Fourier transform with respect to the relative time
	$t_{rel} = t - t'$ at a given average time
	$t_{av}=(t+t')/2$~\cite{RevModPhys.86.779}. The integrand is multiplied by
	a Hann window in $t_{rel}$, which suppresses
	the spectral leakage due to the finite integration range. Defining the time-dependent LDOS as
	$\tilde{\mathcal{N}}_{N}(t_{av},\omega)=\Im\{\tilde{\mathcal{T}}_{N}^{<}(t_{av},\omega)\}/2\pi$,
	its variation induced by the pulse through $\Delta$ is given by $\delta\tilde{\mathcal{N}}_N = \tilde{\mathcal{N}}_N^{w.p.} -
	\tilde{\mathcal{N}}_N^{n.p.}$.
	
	The resulting differential map is reported in Fig.~\ref{fig:fig11}(b), where
	positive (red) and negative (blue) regions denote, respectively, a
	pulse-induced increase and decrease in the occupation of the levels at
	energy $\omega$ and time $t_{av}$. The variation is confined to a time window $T_{env}$ corresponding to the
	transit of the pulse envelope ($t_\alpha=5\,\mathrm{fs} \leq t_{av} \leq t_\beta=
	25\,\mathrm{fs}$). Within it, $\delta\tilde{\mathcal{N}}_{N}$ consists of
	alternating lobes whose period in $t_{av}$ is $\simeq 2$~fs, that is,
	$T_{per}$ rather than $T_{per}/2$. The
	response therefore follows the electric field $\mathcal{E}(t)$ itself, and
	not its intensity $\mathcal{E}^2(t)$: the electrons at the tip apex do not simply undergo
	passive thermal heating, but experience a transient polarization governed by the optical field. Moreover, this coherent oscillation spans the whole occupied bandwidth,
	showing that the field acts on the entire electron distribution at the apex and not only on the states closest to the Fermi level~\cite{PhysRevB.88.195434}.
	
	The oscillating pattern of Fig.~\ref{fig:fig11}(b) averages out over an
	optical cycle, and is therefore not directly reflected in a time-integrated
	quantity such as the far-field distribution (see
	Eqs.~\eqref{eq:full_G_Jdelta}--\eqref{eq:n_eps}). To single out the effect left by the pulse on the emitter, we average
	$\delta\tilde{\mathcal{N}}_N$ over the transit of the envelope,
	\begin{equation}
		\label{eq:delta_bar}
		\mathcal{D}_{N}\left(\omega\right)=\frac{1}{T_{env}}
		\int_{t_\alpha}^{t_\beta} dt_{av}\,\delta\tilde{\mathcal{N}}_{N}\left(t_{av},\omega\right),
	\end{equation}
	which is reported in Fig.~\ref{fig:fig11}(c). Averaging cancels the coherent oscillations discussed above and leaves a net variation of the occupation, dominated by two features: a loss of occupation of the levels just below the Fermi energy,
	and a gain of occupation of those lying deeper in the band, between $-2$
	and $-1$~eV. These features can be put in correspondence with the emission
	profile of Fig.~\ref{fig:fig11}(a). The loss of occupation below the Fermi
	energy involves the levels from which the multi-photon peaks originate, and
	is reflected in their reduction. On the other hand, the gain deeper in the occupied band
	involves levels that emit at correspondingly lower energies, and is
	reflected both in the leftward broadening of the photonic sidebands and in
	the filling of the valleys separating consecutive photon orders. Finally, two weaker structures also emerge in $\mathcal{D}_N$: a loss of
	occupation below the lower edge of the non-interacting dispersion
	($\omega=-2\,$eV), in the energy range already reshaped by correlations at
	equilibrium, and a gain above the Fermi energy, which marks the onset of a
	pulse-induced hot-electron tail at the tip apex.
	
	\section{Summary and conclusions}\label{sec:Summary-and-Conclusions}
	
	In this work, we have presented a theoretical and numerical framework based on the non-equilibrium Green's function formalism to simulate ultrafast laser-assisted electron emission from a metallic nanotip. By modeling the system through a one-dimensional tight-binding chain coupled to a vacuum region with a time-oscillating barrier profile, we have evaluated the far-field energy distribution of the emitted electrons in different cases.
	
	First, we have considered the uncorrelated limit, which serves as our benchmark, by treating the tunneling process perturbatively. We have inferred that the emission dynamics can be interpreted as the absorption of photons by electrons within the tip. The interplay between different emission regimes (photofield emission, multi-photon, and above-threshold photoemission) has been captured and explained in terms of the Fourier amplitudes of the modified vacuum propagators. In addition, we have provided an illustrative picture of how external parameters, such as the pulse and barrier field intensities, govern electron emission.
	
	Afterward, we have introduced weak electron-electron interactions via the self-consistent local second-Born approximation. By considering the effects of correlations on the local density of occupied states, we have shown that these interaction-induced modifications are transferred to the far-field energy distribution. In particular, the emitted electrons carry the signature of these many-body interactions, behaving as quasiparticles ``dressed'' by correlations even far from the emitting tip.
	
	Finally, by self-consistently solving the Dyson equation at the
	matter-vacuum interface, we have gone beyond the perturbative limit to
	capture the laser pulse's dynamic back-action on the emitter. Coupling to
	the driven vacuum modulates in time the occupation of the levels at the tip apex
	with the optical period, so that the electrons are periodically polarized
	rather than passively heated. When averaged over the pulse, this modulation
	leaves a net redistribution of the occupation, including the onset of a
	hot-electron tail. Compared to the regime without back-action, these
	effects manifest in the far-field energy spectrum through a reduction of the multi-photon peaks, a leftward broadening of the sidebands, and a partial filling of the valleys.
	
	The proposed theoretical framework presents several advantages over previous phenomenological models. Because single- and multi-photon transitions emerge naturally in our approach, it circumvents the need for \textit{ad hoc} assumptions about the photon-excited many-body state. Moreover, by including the emitting material in the simulation, our framework establishes a direct mapping between the emitter's local density of occupied states and the measurable far-field energy distribution. Looking forward,
	this opens several pathways for future research: for instance, extending the treatment of electronic interactions to the strongly correlated regime would enable the exploration of how Mott-insulating features manifest in the emission spectrum. Additionally, incorporating the optical pulse's spatial penetration into the bulk would provide further insight into how transient bulk excitations can affect the emission process. Finally, the unconstrained tunability of the laser parameters makes this approach suited to investigate other laser-driven phenomena, such as electron-matter rescattering, thereby offering a robust tool for probing out-of-equilibrium electron emission from quantum materials.
	
	\section*{Acknowledgments}
	
	L.J.D'O. and A.A. acknowledge support from MIUR under Project No. PRIN 2017RKWTMY.
	L.J.D'O. acknowledges support from the Erasmus+ Traineeship program and the PNRR MUR project PE0000023-NQSTI.
	Funding is acknowledged from the Deutsche Forschungsgemeinschaft (DFG) via the Cluster of Excellence CUI: Advanced Imaging of Matter -- EXC 2056 -- project ID 390715994.
	
	\appendix
	\section{Determining the propagators along the vacuum chain\label{app:A}}
	
	To implement the algorithm described in Sec.~\ref{sec:Model-and-Method}, we first need to determine the isolated vacuum propagators. Because the isolated vacuum chain is non-interacting, solving the equation of motion for the NEGFs along it is straightforward.
	
	The Hamiltonian for the isolated vacuum chain, mapped onto a tight-binding grid of $M$ sites, reads
	\begin{equation}
		\hat{\mathcal{H}}_{v}(t)=\sum_{i,j=1}^{M}\hat{\bar{c}}_{i}^{\dagger}h_{i,j}^{v}(t)\hat{\bar{c}}_{j},
	\end{equation}
	where $\hat{\bar{c}}_{i} = \hat{c}_{i+N}$ and $h_{i,j}^{v}(t)$ represents the single-particle Hamiltonian matrix for the vacuum space. The contour-ordered Green's function in the isolated vacuum is defined as $g_{i,j}(t,t') = -i \langle \mathcal{T}_{\mathcal{C}} \hat{\bar{c}}_{i}(t) \hat{\bar{c}}_{j}^{\dagger}(t') \rangle$. Its time evolution is ruled by the single-particle time-evolution (SPTE) operator $\hat{u}(t,t') = \mathcal{T}_{\mathcal{C}} \exp[-i \int_{t'}^t d\bar{t}\, h^v(\bar{t})]$.
	
	As we will prove in Appendix~\ref{app:B}, computing far-field observables requires only the retarded component of the propagator, $g^R_{i,j}(t,t')$. It is given by
	\begin{equation}
		g_{i,j}^{R}(t,t') = -i \theta(t-t') \hat{u}_{i,j}(t,t').
	\end{equation}
	Specifically, since only the retarded Green's function at the first vacuum site is required, we can focus on the first column ($j=1$) of the SPTE operator. Therefore, the equation of motion for $\hat{u}$ to be solved reduces to a first-order linear differential equation coupling exclusively the elements of its first column:
	\begin{equation}
		i\partial_{t} \hat{u}_{i,1}(t,t') = \sum_{l=1}^{M} h_{i,l}^{v}(t) \hat{u}_{l,1}(t,t'),
	\end{equation}
	with the initial condition $\hat{u}_{i,1}(t,t) = \delta_{i,1}$.
	
	This equation can be solved numerically by defining an $n \times n$ time grid with step size $h_{\mathrm{step}}$, according to the following procedure:
	\begin{enumerate}
		\item Fix $t'=0$ and integrate the differential equation forward for $t \ge 0$, obtaining the vector $\hat{u}_{i,1}(t,0)$ up to the maximum time $t_{max}$.
		\item Increment the initial time to $t'=h_{\mathrm{step}}$ and solve the system for $t \ge h_{\mathrm{step}}$, yielding $\hat{u}_{i,1}(t,h_{\mathrm{step}})$.
		\item Repeat this process for each value of $t'$ on the time grid up to $t'=t_{max}$.
	\end{enumerate}
	In this way, the retarded propagator $g^R_{i,1}(t,t')$ can be reconstructed for $t \ge t'$. Subsequently, it is used to evaluate the hybridization function and the far-field energy distribution.
	
	\section{The imaginary-time components\label{app:B}}
	
	To evaluate the far-field energy distribution, the lesser component of the full Green's function, $\mathbb{G}^<$, must be computed. Following the Langreth rules for analytic continuation on the Kadanoff-Baym contour, the lesser component of $\mathbb{G}_{i,j}$ evaluated in the vacuum region expands as follows~\cite{RevModPhys.86.779, Eckstein2018, Stefanucci-Leeuwen:2013}:
	\begin{align}
		\mathbb{G}_{i,j}^{<} &= g_{i,j}^{<} + (g_{i,1} \ast \bar{\mathcal{T}}_{N} \ast g_{1,j})^{<} \nonumber \\
		&= g_{i,j}^{<} + g_{i,1}^{R} \cdot \bar{\mathcal{T}}_{N}^{R} \cdot g_{1,j}^{<} + g_{i,1}^{R} \cdot \bar{\mathcal{T}}_{N}^{<} \cdot g_{1,j}^{A} \nonumber \\
		&\quad + g_{i,1}^{<} \cdot \bar{\mathcal{T}}_{N}^{A} \cdot g_{1,j}^{A} + g_{i,1}^{tv} \star \bar{\mathcal{T}}_{N}^{vt} \cdot g_{1,j}^{A} \nonumber \\
		&\quad + g_{i,1}^{R} \cdot \bar{\mathcal{T}}_{N}^{tv} \star g_{1,j}^{vt} + g_{i,1}^{tv} \star \bar{\mathcal{T}}_{N}^{M} \star g_{1,j}^{vt},
	\end{align}
	where $\cdot$ and $\star$ denote the real-time and imaginary-time convolution, respectively, and $\bar{\mathcal{T}}_{N}(t,t')=t_{c}(t)\,\mathcal{T}_{N}(t,t')\,t_{c}^{*}(t')$ is the lattice counterpart of the emitter term introduced in Sec.~\ref{sec:adi}, obtained by replacing $J_0$ with the matter-vacuum hopping amplitude $t_c$.
	
	At equilibrium ($t=0$), we assume that electrons are confined within the tip and the vacuum is empty, that is, $g^< \sim 0$. Furthermore, because the initial state is adiabatically prepared from the equilibrium one, the coupling is set to zero for $t \le 0$, so that $t_c$ vanishes identically along the imaginary-time branch of the contour. As a consequence, all the components of $\bar{\mathcal{T}}_{N}$ with at least one time argument on that branch (i.e., the Matsubara $M$, left-mixing $tv$, and right-mixing $vt$ components) vanish: $\bar{\mathcal{T}}_{N}^{M} = \bar{\mathcal{T}}_{N}^{tv} = \bar{\mathcal{T}}_{N}^{vt} = 0$.
	It follows that all imaginary-time convolutions drop out of the expression for $\mathbb{G}^<$, simplifying it to a compact form restricted exclusively to the real-time domain:
	\begin{equation}
		\mathbb{G}_{i,j}^{<}(t,t') = \int_{0}^{t_{max}} dt_1 dt_2 \, g_{i,1}^{R}(t,t_1) \bar{\mathcal{T}}_{N}^{<}(t_1,t_2) g_{1,j}^{A}(t_2,t').
	\end{equation}
	This occurrence proves that, by exploiting adiabatic initialization, the computationally demanding Matsubara and mixing components of the vacuum NEGFs can be safely bypassed. Thus, since $g_{i,j}^{A}\left(t,t'\right)=\left[g_{j,i}^{R}\left(t',t\right)\right]^{*}$, only the real-time retarded components of the isolated vacuum propagators are required to reconstruct the far-field dynamics and associated observable features.

\end{document}